\documentclass[12pt,times]{article}
 \usepackage[pdftex]{color,graphicx}
\usepackage{enumerate}
\usepackage{amsmath}
\usepackage{amssymb}
\usepackage{alltt}
\usepackage{setspace}
\usepackage{subfigure}
\usepackage{sidecap}
\usepackage{sectsty}
\partfont{\large}
\sectionfont{\large}
\subsectionfont{\small}
\usepackage[left=2cm,top=2cm,right=3cm,head=1cm,foot=1cm]{geometry}
\usepackage{mathrsfs}
\numberwithin{equation}{section}
\let\oldsqrt\sqrt
\def\sqrt{\mathpalette\DHLhksqrt}
\def\DHLhksqrt#1#2{%
\setbox0=\hbox{$#1\oldsqrt{#2\,}$}\dimen0=\ht0
\advance\dimen0-0.2\ht0
\setbox2=\hbox{\vrule height\ht0 depth -\dimen0}%
{\box0\lower0.4pt\box2}}
\newcommand{\al}{\alpha}
\newcommand{\g}{\gamma}
\newcommand{\e}{\varepsilon}
\newcommand{\ta}{\theta}
\newcommand{\z}{\zeta}

\newcommand{\ph}{\varphi}
\newcommand{\da}{\dagger}

\newcommand{\la}{\mathscr{L}}
\newcommand{\ld}{\lambda}
\newcommand{\Ld}{\Lambda}

\newcommand{\M}{\mathcal M}

\newcommand{\+}{\oplus}
\newcommand{\x}{\times}

\newcommand{\rx}{\rtimes}
\newcommand{\ml}{\left(\begin{matrix}}
\newcommand{\mr}{\end{matrix}\right)}

\newcommand{\U}{\mathcal U}

\newcommand{\w}{\omega}
\newcommand{\W}{\Omega}

\newcommand{\bra}{\langle}
\newcommand{\ket}{\rangle}
\newcommand{\tr}{\text{tr}}
\newcommand{\op}{\mathcal O}
\newcommand{\del}{\delta}
\newcommand{\Del}{\Delta}
\newcommand{\lrarrow}{\leftrightarrow}
\newcommand{\zb}{\mathbb Z}

\newcommand{\ep}{\epsilon}
\newcommand{\re}{\text{Re}}

\newcommand{\half}{\tfrac{1}{2}}
\newcommand{\third}{\tfrac{1}{3}}
\newcommand{\fourth}{\tfrac{1}{4}}

\newcommand{\s}{\sigma}

\newcommand{\msol}{\Del m^2_{\text{sol}}}
\newcommand{\matm}{\Del m^2_{\text{atm}}}

\newcommand{\sal}{\!\!\!\sum_{\al\,=\,e,\mu,\tau}\!\!\!}
\newcommand{\delCP}{\del_{\text{CP}}}
\newcommand{\sig}{\Sigma(81)}

\newcommand{\diag}{\text{diag}}

\newcommand{\eff}{\text{eff}}

\newcommand{\ebar}{\overline e}
\newcommand{\mubar}{\overline\mu}
\newcommand{\taubar}{\overline\tau}

\begin{document}
\title{\Large \textbf{Lepton Private Higgs and the discrete group $\Sigma(81)$}}
\author{Yoni BenTov$^{1}$ \and A. Zee$^{1,2}$}
\date{}
\maketitle
\begin{flushleft}
\textit{$^1$\,Department of Physics, University of California, Santa Barbara CA 93106}\\
\textit{$^2$\,Kavli Institute for Theoretical Physics, University of California, Santa Barbara CA 93106}
\end{flushleft}
\begin{abstract}
We use the discrete group $\sig \equiv (\zb_3\times\zb_3\times\zb_3)\rx\zb_3$ to explore a particular region of parameter space in the Private Higgs model. In doing so we suggest a relation among the off-diagonal entries of the neutrino mass matrix and a possible explanation for the muon magnetic moment anomaly, $a_\mu^{\text{exp}}-a_\mu^{\text{SM}} \sim 10^{-9}$. We predict three new nearly degenerate Higgs doublets with masses of order $\sim 500$ GeV to $\sim 1$ TeV, and three nearly degenerate SM-singlet TeV-scale neutrinos. The largest scale in the model is $\sim 10$ TeV, so there is no severe hierarchy problem. The appendix is devoted to the group theory of $\sig$.
\end{abstract}
\section{Introduction}\label{sec:intro}
The matter content of the Standard Model (SM) of particle physics comes in three generations, which are identical except for their widely disparate mass scales. At sufficiently high energies\footnote{For the quarks, ``sufficiently high energy" means larger than the top mass $m_t \approx 173$ GeV, while for the leptons the required scale is much lower, given by the $\tau$ mass $m_\tau \approx 1.78$ GeV.} all particles become effectively massless, and so it is possible that a high energy completion of the SM is symmetric under permutation of the generation labels \cite{generalizedpermutation}. If we insist that the fermions acquire mass perturbatively from the Higgs mechanism while having Yukawa couplings of comparable magnitudes, then we must extend the SM to include multiple Higgs doublets, which are to be permuted along with the fermion generations.
\\\\
The challenge for all models of this type is how to implement a permutation symmetry that can accommodate the relations $m_d \ll m_s\ll m_b$ and $m_u\ll m_c \ll m_t$ for the quarks, and $m_e \ll m_\mu \ll m_\tau$ for the charged leptons, while at the same time allowing for nonzero mixing angles in the CKM and PMNS matrices.
\\\\
Among the myriad theoretical possibilities, we will take motivation from one particular class of multi-Higgs models, the models of ``Private Higgs"-type \cite{PH, leptonPH}, and extend the idea to include a permutation symmetry. This will immediately suggest a particular non-abelian discrete flavor group, $\sig \equiv (\zb_3\times\zb_3\times\zb_3)\rx\zb_3$, which was first applied to leptons by E. Ma \cite{sig81ma} and later studied by other authors \cite{sig81other}.
\\\\
The Private Higgs (PH) philosophy is to introduce a Higgs doublet $\phi_q$ for each quark $q = d,s,b,u,c,t$ and a Higgs doublet $\phi_\al$ for each charged lepton $\al = e,\mu,\tau$. The fermions are supposed to obtain masses from the Yukawa interactions\footnote{We use two-component spinor notation for fermions. Here $q_a = (u_a,d_a) \sim (3,2,+\tfrac{1}{6})$ are the left-handed quark doublets, $\overline d_a \sim (3^*,1,+\third)$ and $\overline u_a \sim (3^*,1,-\tfrac{2}{3})$ are the left-handed antiquark singlets, $\ell_\al = (\nu_\al,e_\al) \sim (1,2,-\half)$ are the left-handed lepton doublets, and $\overline e_\al \sim (1,1,+1)$ are the left-handed antilepton singlets. The fields $\phi_{D_a} = (\phi_{D_a}^0,\phi_{D_a}^-) \sim (1,2,-\half)$, $\phi_{U_a} = (\phi_{U_a}^+,\phi_{U_a}^0) \sim (1,2,+\half)$, and $\phi_\al = (\phi_\al^0,\phi_\al^-) \sim (1,2,-\half)$ are Higgs doublets. Note that $\phi_{D_a}$ and $\phi_\al$ have hypercharge $Y=-1$, while $\phi_{U_a}$ has hypercharge $Y = +1$. The dot denotes the $SU(2)$-invariant antisymmetric product, and the subscripts are understood as $\overline d_1 \equiv \overline d$, $\phi_{D_1} \equiv \phi_d$, and so on.}
\begin{equation}\label{eq:PHyukawa}
\la_{\text{Yuk}} = \!\!\!\sum_{a\,=\,1,2,3}\!\!\left(y_{D_a}q_a\!\cdot\phi_{D_a}\overline d_a -y_{U_a}q_a\!\cdot\!\phi_{U_a}\overline u_a\right)+\sal y_{\al}\,\ell_\al\!\cdot\phi_{\al}\, \ebar_\al+h.c.
\end{equation}
The mass of the each fermion is then determined by the vacuum expectation value (VEV) of its associated Higgs\footnote{The idea that the fermion masses arise from a small VEV of a second Higgs doublet is an old idea \cite{haber kane sterling, grimus lavoura radovcic}. A different model with one Higgs for each lepton flavor was proposed by Grimus and Lavoura \cite{grimus1}.} doublet:
\begin{equation}
m_{d,s,b} = y_{d,s,b}\bra\phi^0_{d,s,b}\ket\;,\;\; m_{u,c,t} = y_{u,c,t} \bra\phi^0_{u,c,t}\ket\;,\;\; m_{e,\mu,\tau} = y_{e,\mu,\tau}\bra\phi^0_{e,\mu,\tau}\ket\;.
\end{equation}
To ensure that each Higgs serves only its associated fermion in Eq.~(\ref{eq:PHyukawa}), we need to impose a family-dependent discrete symmetry \cite{discretefamilysymmetry}. For simplicity and concreteness let us first focus on the charged leptons. Observing that each Yukawa interaction is a product of three fields, we impose the symmetry\footnote{This differs from the $\zb_2$ symmetries used in the original Private Higgs model.}
\begin{equation}\label{eq:Z3leptonsymmetries}
\zb_3^e\times\zb_3^\mu\times\zb_3^\tau\;.
\end{equation}
The group $\zb_3^e$ multiplies the fields $\ell_e \equiv (\nu_e,e)$, $\phi_e \equiv (\phi_e^0,\phi_e^-)$, and $\overline e_e \equiv \ebar$ by the phase $\w \equiv e^{\,i2\pi/3}$, and leaves the other fields alone. The other $\zb_3^\al$ for $\al = \mu,\tau$ are defined analogously. 
\\\\
We would now like to suppose that the three lepton generations are interchangeable above $m_\tau$ and therefore impose a $\zb_3^C$ symmetry that cycles the fields as $(\ell_e\phi_e\ebar, \ell_\mu\phi_\mu\mubar, \ell_\tau\phi_\tau\taubar)$ $\to$ $(\ell_\tau\phi_\tau\taubar,\ell_e\phi_e\ebar,\ell_\mu\phi_\mu\mubar)$ $\to$ $(\ell_\mu\phi_\mu\mubar,\ell_\tau\phi_\tau\taubar,\ell_e\phi_e\ebar)$. Since the label $\al = e,\mu,\tau$ is now supposed to denote a triplet representation rather than a collection of singlets, the transformations of Eq.~(\ref{eq:Z3leptonsymmetries}) do not commute with those of $\zb_3^C$ \cite{grimus2}.
\\\\
Thus we are led quite naturally to consider the non-abelian discrete group
\begin{equation}
\sig \equiv (\zb_3^e\times\zb_3^\mu\times\zb_3^\tau)\rx \zb_3^C\;.
\end{equation}
The leptons and lepton-Higgs fields are assigned to the defining triplet representation 3, which is complex, and the product $3\x3\x3$ contains an invariant singlet. The group $\sig$ has three other complex triplets, so we can incorporate the quarks into this framework by simply assigning the quarks and quark-Higgs fields to a triplet representation distinct from the defining 3. Later on, we will find that we must also impose an additional abelian symmetry to forbid certain bare terms in the scalar potential [see the discussion above Eq.~(\ref{eq:symmetrybreakingsequence})]; this is a common ingredient in many flavor models based on non-abelian groups.
\\\\
We emphasize at this point that our motivating philosophy is to provide an existence proof that the Private Higgs mechanism is compatible with non-abelian discrete flavor symmetries by providing an explicit example based on the group $\sig \times X$ (where $X$ is abelian), not to predict a particular form for the neutrino mixing matrix. This work is also an example of exploring a different regime of parameter space in the Private Higgs framework that differs from the original proposal in \cite{PH, leptonPH}.
\\\\
We should also point out that despite the form of the high-energy Lagrangian, our model is \textit{not} fermiophobic at low energies \cite{PH at LHC} and is compatible with the celebrated recent discovery of an SM-like Higgs state with mass $125$ GeV \cite{CMS, ATLAS}. The reason is as follows (consider the electron flavor for concreteness). The Lagrangian for the Private Higgs field $\phi_e$ is of the form
\begin{equation}
\la_{\phi_e} = \phi_e^\da (D^\da D-\hat M_{\phi_e}^2) \phi_e - (J_e^\da \phi_e + h.c.)
\end{equation}
where\footnote{For simplicity we drop terms with a nontrivial $SU(2)$ index structure in the Higgs potential, which play no essential role in the present argument.} $\hat M_{\phi_e}^2 = M_{\phi_e}^2 +$ (quartic couplings to Higgs fields and SM-singlet scalars) is taken \textit{positive}, and $J_e$ includes a cubic coupling to the SM-like ``$W$-Higgs" [see Eq.~(\ref{eq:W higgs})] and SM-singlet scalar $S_e$, as well as the electron Yukawa interaction:
\begin{equation}
(J_e)_i = \mu_e (\phi_W^\da)^j\e_{ji}+y_e\ml \nu_e\\ e \mr_i \bar e\;.
\end{equation}
Upon integrating out the $\phi_e$ field, we obtain an effective interaction between the SM-like Higgs and the electron:
\begin{equation}
\la_\eff = -y_e \frac{\mu_e^*\bra S_e\ket^*}{\hat M_{\phi_e}^2}(\phi_W)_i\e^{ij}\ml \nu_e\\ e \mr_j \bar e + h.c. +...
\end{equation}
Thus, to leading order, the $h e\bar e$ coupling in this model is $(m_e/v)(1+...)$, where $v^2 = 2(|\bra\phi_W^0\ket|^2+...) = (246\,\text{GeV})^2$, which is the same as in the SM up to corrections that are ``generically" $\op(v^2/\M^2) \sim 1\%$ if $\M \sim$ TeV, but which could in principle be substantial depending on the values of the various parameters in the scalar potential. We refer to \cite{PH at LHC} for a more detailed treatment, in which the LHC phenomenology was studied explicitly for a general class of ``Private Higgs"-type models.
\\\\
The rest of this paper is organized as follows. In Section~\ref{sec:leptons} we discuss the charged lepton masses and a TeV-seesaw mechanism for neutrino masses under the assumption of $\sig$ symmetry. In Sections~\ref{sec:numixing} and~\ref{sec:data} we discuss neutrino mixing and predict a relation among the off-diagonal entries of the Majorana neutrino mass matrix. In Section~\ref{sec:g-2} we discuss the anomalous magnetic moment of the muon. In Section~\ref{sec:LFV} we discuss constraints due to lepton flavor violation. In Section~\ref{sec:quarks} we comment briefly on quark masses, but leave the construction of a realistic CKM matrix for future work. In Sections~\ref{sec:potential} and~\ref{sec:soft} we discuss the scalar potential of the model and justify the symmetries and scales that are assumed for the phenomenology of the previous sections. In Section~\ref{sec:discussion} we summarize our results and suggest directions for future work. In the appendix we discuss the group theory of $\sig$.
\section{Lepton masses and TeV-scale seesaw}\label{sec:leptons}
Let the lepton doublets\footnote{Since we are dealing with leptons, which are all color singlets, we suppress the $SU(3)$ quantum number.} $\ell = (\nu,e) \sim (2,-\half)$ and the antilepton singlets $\ebar \sim (1,+1)$ transform as the defining 3 representation of $\sig$. As discussed in the introduction, we introduce a collection of three $SU(2)\times U(1)$ ``lepton-Higgs" doublets $\phi_e,\phi_\mu,\phi_\tau$, collectively denoted by the field $\phi_\ell = (\phi_\ell^0,\phi_\ell^-) \sim (2,-\half)$, which also transforms as a 3 under $\sig$. Denoting the $\sig$ components by the flavors $e,\mu,\tau$, we have the $\sig$-invariant Yukawa interactions
\begin{equation}\label{eq:lepton Yukawa interactions}
\la_{\text{Yuk}}^\ell = y_\ell\left( \ell_e\!\cdot\!\phi_e\,\overline e+\ell_\mu\!\cdot\!\phi_\mu\,\overline\mu+\ell_\tau\!\cdot\!\phi_\tau\,\overline\tau\right)+h.c.
\end{equation}
Thus upon writing $\phi_\al^0 = \tfrac{1}{\sqrt2}v_\al\,e^{\,i\ta_\al}+...$ and rephasing the charged lepton fields, the charged leptons obtain masses
\begin{equation}
m_\al = \tfrac{1}{\sqrt2}y_\ell v_\al\;.
\end{equation}
Given the common Yukawa coupling across the three generations, we have
\begin{equation}
m_e:m_\mu : m_\tau = v_e:v_\mu : v_\tau\;.
\end{equation}
Thus with $y_\ell \sim 1$ the charged lepton masses are determined by the vacuum expectation values of their associated lepton-Higgs fields.
\\\\
Before proceeding further, we should comment that the coupling $y_\ell$ depends on energy scale $\mu$, and is defined in Eq.~(\ref{eq:lepton Yukawa interactions}) at a $\sig$-invariant scale $\M$. To compute observables at low energy, we should use the renormalization group (RG) to find that the flavor-independent coupling $y_\ell(\mu)$ splits into three distinct effective couplings, $y_e^{\text{eff}}(\mu)$, $y_\mu^{\text{eff}}(\mu)$, and $y_\tau^{\text{eff}}(\mu)$, at scales $\mu \ll m_e \ll m_\mu \ll m_\tau$. In practice, since there are no superheavy scales in our model, these effects are small\footnote{
Since flavor changing interactions are small, the dominant loop correction to the $\ell_\al \phi_\al \bar e_\al$ vertex comes solely from the lepton $\al$ circulating in the loop. For large Yukawa couplings, the gauge contributions are subleading, and we estimate the 1-loop beta function as $\mu\, dy_\ell/d\mu \approx a\,y_\ell^3/(4\pi)^2$, with $a > 0$.
The Yukawa couplings remain independent of lepton flavor at low energy up to corrections of $O(10^{-2})$ if $y_\ell(\M) \approx 1$. For smaller values of $y_\ell(\M)$, the running is negligible.} even for Yukawa couplings $y_\ell(\M)\approx 1$.
\\\\
Let us now introduce another Higgs doublet,
\begin{equation}\label{eq:W higgs}
\phi_W = (\phi_W^0,\phi_W^-) \sim (2,-\half)\;,
\end{equation}
which does not transform under any flavor symmetry. This $W$-Higgs provides the dominant contribution to the mass of the $W^\pm$ bosons:
\begin{equation}\label{eq:mW}
m_W = \half g v_W\left[1+O\!\left(\frac{v_{\text{other}}^2}{v_W^2}\right)\right]
\end{equation}
where $v_{\text{other}}$ denotes the contributions of the non-$W$ Higgs VEVs. The $Z$ obtains a mass $m_Z = m_W/\cos\ta_W$, just as in the SM. So far this is the same as in the original Private Higgs model for leptons\footnote{In that model the role of $\phi_W$ is played by a top-Higgs $\phi_t$.}, but with the additional constraint $y_e = y_\mu = y_\tau \equiv y_\ell$.
\\\\
For the neutrino sector, introduce an SM-singlet antineutrino field $N \sim (1,0)$ and another Higgs doublet $\phi_\nu = (\phi_\nu^+,\phi_\nu^0) \sim (2,+\half)$, each of which is also to transform as a 3 under $\sig$. Just as for the charged leptons, the $\sig$-invariant Yukawa interactions for the neutrinos are diagonal in flavor and have a flavor-independent coupling:
\begin{equation}
\la_{\text{Yuk}}^\nu = -y_\nu\left( \ell_e\!\cdot\!\phi_{\nu_e}N_e+\ell_\mu\!\cdot\!\phi_{\nu_\mu}N_\mu+\ell_\tau\!\cdot\!\phi_{\nu_\tau}N_\tau\right)+h.c.
\end{equation}
As a result of the opposite hypercharge assignments for $\phi_\ell$ and $\phi_\nu$, only the former gives masses to the charged leptons while only the latter provides Dirac masses for the neutrinos. 
\\\\
At this stage a bare mass for the gauge-singlet neutrinos $N_\al$ is forbidden by the flavor symmetry. The Private Higgs model \cite{PH, leptonPH}, on which the present model is based, requires the existence of SM-singlet scalars to enable the heavy Higgs fields $\phi_\al$ and $\phi_{\nu_\al}$ to have large masses but small VEVs [see Eq.~(\ref{eq:privatehiggsmass})]. Therefore, we require the addition of SM-singlet complex scalar fields $S_\ell \sim (1,0)$, which transform as a triplet under $\sig$. Fortuitously, these are precisely the degrees of freedom required to generate seesaw masses for the gauge-singlet neutrinos. 
\\\\
The $\sig$-invariant Yukawa interaction
\begin{equation}
\la_{\text{Yuk}}^N = -\half y_N(S_eN_eN_e+S_\mu N_\mu N_\mu+S_\tau N_\tau N_\tau)+h.c.
\end{equation}
leads to flavor-diagonal Majorana masses for the gauge-singlet neutrinos:
\begin{equation}
M_{N_\al} = y_N\bra S_\al\ket\;.
\end{equation}
Upon integrating out the $N_\al$, the light neutrinos will obtain a flavor-diagonal Majorana mass matrix from the usual seesaw mechanism:
\begin{equation}
(m_\nu^{\text{seesaw}})_{\al\beta} = -\,\frac{(y_\nu\bra\phi_{\nu_\al}^0\ket)^2}{y_N\bra S_\al\ket}\;\del_{\al\beta}\;.
\end{equation}
Here the VEVs $\bra S_\al\ket$ set the seesaw scale, which in our model will be $\sim$ TeV \cite{TeVseesaw}. The neutrino mass scale $m_\nu \lesssim$ eV can be obtained by taking the neutrino-Higgs VEVs $\bra\phi_{\nu_\al}^0\ket$ to be roughly of order the electron mass \cite{TeVseesawelectronmass}. Just as for the charged-lepton-Higgs fields, we define $\bra\phi_{\nu_\al}^0\ket \equiv \tfrac{1}{\sqrt2}v'_\al\,e^{\,i\ta'_\al}$ and work with the manifestly real parameters $v'_\al$.
\\\\
At this stage the neutrinos do not oscillate, so we still need to generate nonzero off-diagonal contributions to the neutrino mass matrix.\footnote{This construction is similar in spirit to an early model of Fukugita and Yanagida \cite{nooscillation}. Interestingly, at that time the lack of oscillations in the model was considered a virtue rather than a deficiency.} 
\section{Neutrino mixing}\label{sec:numixing}
\begin{figure}[h]
\begin{center}
\fbox{
	\begin{minipage}{16 cm}
	\centering
	\includegraphics[width=90mm]{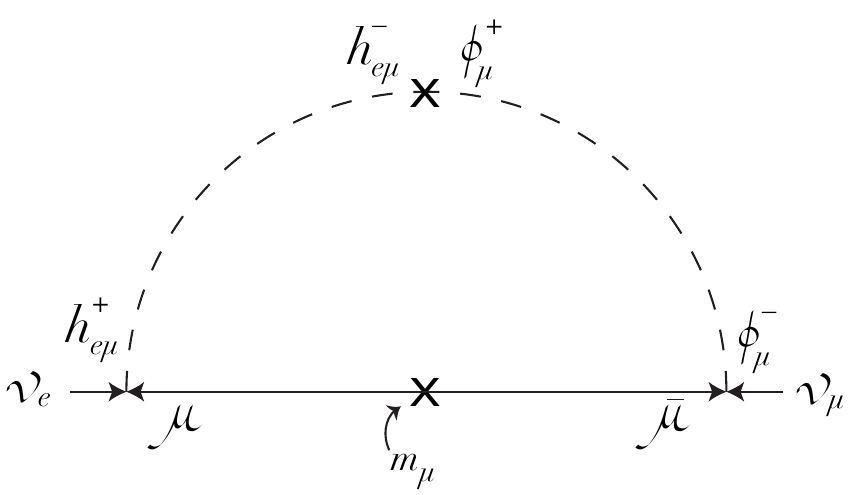}
	\caption{\small{One-loop contribution to $(m_\nu)_{e\mu}$. The diagrams are labeled by the fields that participate in each interaction vertex, and the arrows denote fermion spin. The lower X is to remind the reader that the $\mu\bar\mu$ propagator, which conserves chirality, includes a factor of $m_\mu$ in the numerator. The upper X is to indicate that we are working in an approximate basis in which the leading order scalar mass squared terms are large compared to the post-symmetry breaking corrections from the potential, and the leading order contribution to $(m_\nu)_{e\mu}$ thereby contains two scalar propagators and one quadratic mixing term that changes the $h_{e\mu}^\pm$ line to a $\phi_\mu^\pm$ line. There is also an analogous contribution with the electron running in the loop, which has a factor of $m_e$ in the numerator.}}
		\label{fig:diagram1}
	\end{minipage}
}
\end{center}
\end{figure}

To generate nonzero mixing angles for the light neutrinos, we introduce a flavor triplet of singly charged spin-0 bosons, $h^-$, whose interactions result in off-diagonal entries for the neutrino mass matrix at one loop \cite{zeemodel}. 
\\\\
In contrast to the other fields of the model, we assign $h^-$ to a different triplet representation, $3'$, and thereby denote its components by\footnote{As explained in the appendix, the labeling results from forming the $3'$ by decomposing the product $3\x3$ into irreducible representations. See Eqs.~(\ref{eq:(z_e,I)on3'}) and~(\ref{eq:triplet notation}).} $h^- = (h_{\mu\tau}^-,h_{\tau e}^-,h_{e\mu}^-)$. Since $h^+\ell\!\cdot\!\ell \sim (3')^*\x3\x_A3$ contains an invariant [see Eqs.~(\ref{eq:3 x 3 = 3*+3'+3'}) and~(\ref{eq:3x(3')*})], we have a $\sig$-invariant interaction between the charged bosons\footnote{Here $h^+_{\al\beta} \equiv (h^-_{\al\beta})^*$ has electric charge $+1$.} and the lepton doublets:
\begin{equation}\label{eq:hll}
\la_{h\ell\ell} = f\left( h^+_{\mu\tau}\,\ell_\mu\!\cdot\!\ell_\tau+h^+_{\tau e}\,\ell_\tau\!\cdot\!\ell_e+h^+_{e\mu}\,\ell_e\!\cdot\!\ell_\mu\right)+h.c.
\end{equation}
These interactions have a common\footnote{In principle there are flavor-dependent corrections to $f$ at low energy. Since we consider small values $f \lesssim 10^{-2}$ and since the largest scale in the model is $\sim 10$ TeV, these corrections are very small.} coupling $f$, which provides a group theoretic realization of Case A in a study by Ghosal, Koide, and Fusaoka \cite{LFVZdecays}. Note that the $\sig$ symmetry forbids interactions of the form $\ebar Nh^-$.
\\\\
The charged bosons couple to Higgs doublets through three different SM$\times\Sigma(81)$ invariants:
\begin{align}
&t_0 \equiv h^+_{\mu\tau}\phi_\mu\!\cdot\!\phi_\tau+h^+_{\tau e}\phi_\tau\!\cdot\!\phi_e+h^+_{e \mu}\phi_e\!\cdot\!\phi_\mu \label{eq:hphiphi} \\ 
&t_1 \equiv \left(h^+_{\mu\tau} S_\mu \phi_\tau+h^+_{\tau e} S_\tau \phi_e+h^+_{e \mu} S_e \phi_\mu\right)\!\cdot\!\phi_W \label{eq:t1}\\
&t_2 \equiv \left(h^+_{\mu\tau} S_\tau \phi_\mu+h^+_{\tau e} S_e \phi_\tau+h^+_{e \mu} S_\mu \phi_e\right)\!\cdot\!\phi_W \label{eq:t2}
\end{align}
We will see that if the coefficient of the dimension-3 operator $t_0$ is not much larger than $\bra S_\al\ket$, then the contribution to $m_\nu$ from the $t_0$ term will be subleading compared to that from $t_1$ and $t_2$. 
\\\\
Combined with the $h\ell\ell$ interactions of Eq.~(\ref{eq:hll}), the Lagrangian
\begin{equation}\label{eq:hphi}
\la_{h\phi} = \hat M\,t_0+\rho_1\,t_1+\rho_2\,t_2+h.c.
\end{equation}
generates off-diagonal entries in the neutrino mass matrix at one loop.\footnote{These interactions also generate corrections to the diagonal entries of the neutrino mass matrix at two loops \cite{changandzee}. We will drop these as subleading to the diagonal entries generated from the seesaw mechanism.} (Here $\hat M$ is a coupling with dimensions of mass, and $\rho_{1,2}$ are dimensionless couplings.) For example, one contribution to the $(e,\mu)$ entry of $m_\nu$ arises from the diagram in Fig.~\ref{fig:diagram1}. As in the original work \cite{zeemodel}, instead of diagonalizing the large charged scalar mass matrix, we elect to work in the flavor basis and treat the quadratic mixing terms $\sim h^-\phi^+$ as a subleading perturbation. 
\\
\\
For $M_h^2 \gg M_\phi^2$ (where $M_{h}^2$ is the bare coefficient of $h_{e\mu}^+ h_{e\mu}^-$ and $M_{\phi}^2$ is that of $\phi_\mu^+\phi_\mu^-$) the contribution of this diagram to the neutrino mass matrix is
\begin{equation}\label{eq:diagram1}
(m_\nu)_{e\mu}^{\text{diagram 1}} \approx \frac{1}{(4\pi)^2}\frac{fy_\ell\M_{e\mu}^2}{M_h^2}\,m_\mu\ln\frac{M_h^2}{M_\phi^2}\;,\;\;\M_{e\mu}^2 \equiv \rho_1^*\bra S_e\ket^*\bra\phi_W^0\ket^*-\hat M^*\bra\phi_e^0\ket^*\;.
\end{equation}
By the symmetry of the Majorana mass matrix, there is also the same diagram but with the electron and electron-Higgs running in the loop. This gives an additional contribution
\begin{equation}\label{eq:diagram2}
(m_\nu)_{e\mu}^{\text{diagram 2}} \approx \frac{1}{(4\pi)^2}\frac{fy_\ell\M_{\mu e}^2}{M_h^2}\, m_e\ln\frac{M_h^2}{M_\phi^2}\;,\;\; \M_{\mu e}^2 \equiv -\rho_2^*\bra S_\mu\ket^*\bra\phi_W^0\ket^*-\hat M^*\bra\phi_\mu^0\ket^*\;.
\end{equation}
The $(e,\mu)$ entry of the neutrino mass matrix is given by adding the two contributions: $(m_\nu)_{e\mu} = (m_\nu)_{e\mu}^{\text{diagram 1}}+(m_\nu)_{e\mu}^{\text{diagram 2}}$.
\\\\
The relative signs in $\M_{e\mu}^2$ and $\M_{\mu e}^2$ [Eqs.~(\ref{eq:diagram1}) and~(\ref{eq:diagram2}) respectively] can be understood as follows. The interaction $f\,h^+_{e\mu}\ell_e\cdot\ell_\mu = fh^+_{e\mu}(\nu_e\mu-\nu_\mu e)$ in Eq.~(\ref{eq:hll}) results in a vertex $+if$ for diagram 1, but $-if$ for diagram 2. In contrast, the $h^-_{e\mu}\phi_\mu^+$ vertex arising from Eqs.~(\ref{eq:t1}) and~(\ref{eq:hll}) is not related by any symmetry to the $h^-_{e\mu}\phi_e^+$ vertex arising from Eqs.~(\ref{eq:t2}) and~(\ref{eq:hll}), and so they contribute the same way: a factor of $-i\rho_1^*\bra S_e\ket^*\bra\phi_W^0\ket^*$ from the former, and $-i\rho_2^*\bra S_\mu\ket^*\bra\phi_W^0\ket^*$ from the latter. This accounts for the relative sign difference between the contributions from $\rho_1$ and $\rho_2$ to the effective $\M^2_{\al\beta}$ couplings in Eqs.~(\ref{eq:diagram1}) and~(\ref{eq:diagram2}).
\\\\
In addition, the contribution from $\hat M$ appears with the same sign in Eqs.~(\ref{eq:diagram1}) and~(\ref{eq:diagram2}) because the operators $h^+_{e\mu}\phi_e\cdot\phi_\mu$ in Eq.~(\ref{eq:hphiphi}) and $h^+_{e\mu}\ell_e\cdot\ell_\mu$ in Eq.~(\ref{eq:hll}) have identical transformation properties under $SU(2)\times U(1)\times \sig$. In other words, diagram 2 has sign flips relative to diagram 1 in the contributions from both $f$ and $\hat M$, so that the extra signs cancel each other out.
\\\\
Leaving the possibility of CP violation for future work, we assume that all couplings and VEVs are real and write 
\begin{equation}
\bra\phi_\al^0\ket = \tfrac{1}{\sqrt2}v_\al\;,\;\; \bra S_\al\ket = \tfrac{1}{\sqrt2}\widetilde v_\al
\end{equation}
from now on. All ``high-scale" SM-singlet scalar VEVs will be taken essentially equal: $\tilde v_e \approx \tilde v_\mu \approx \tilde v_\tau \equiv \tilde v$ [see Eq.~(\ref{eq:scalar vevs equal})], while the ``low-scale" Higgs VEVs will be taken hierarchical, $v_e \ll v_\mu \ll v_\tau$, and this is obtained by tuning the appropriate low-scale parameters in the scalar potential [see Eq.~(\ref{eq:higgs vevs hierarchical})]. This is the generic approach taken in this model toward dimensionful parameters: those that are $\sim10^2$ GeV or higher are flavor symmetric (up to small corrections), and those that take values below the weak scale are flavor asymmetric. This will be discussed explicitly in Section~\ref{sec:potential}.
\\\\
The vertices from Eqs.~(\ref{eq:t1}) and~(\ref{eq:t2}) are proportional to $\widetilde v_\al v_W$, while those from Eq.~(\ref{eq:hphiphi}) are proportional to $\hat M v_\al \sim \hat M m_\al$, which is suppressed by a power of charged lepton mass. As can be seen from adapting Fig.~\ref{fig:diagram1} to $(m_\nu)_{\mu\tau}$, the only contribution from $\hat M$ that is not obviously small occurs with a factor of $m_\mu m_\tau$ in the numerator. If $\rho_1\sim\rho_2$, then the contribution from $\hat M$ can be dropped provided that $\hat M/\widetilde v \ll 10^2\rho_{1,2}$. With $\rho_{1,2} = O(10^{-1}-1)$, we can take $\hat M$ as large as $\sim (10^{-1}-1)$ TeV. 
\\\\
Dropping the contributions from terms proportional to $\hat M$, and assuming furthermore that all $\widetilde v_\al$ are equal, we arrive at the following off-diagonal entries in the neutrino mass matrix:
\begin{align}
&(m_\nu)_{e\mu} \approx \left(\rho_1\frac{m_\mu}{m_\tau}-\rho_2\frac{m_e}{m_\tau}\right)m_\nu^{\text{1-loop}} \nonumber\\
&(m_\nu)_{\mu\tau} \approx \left(\rho_1-\rho_2\frac{m_\mu}{m_\tau}\right)m_\nu^{\text{1-loop}} \nonumber\\
&(m_\nu)_{\tau e} \approx \left(\rho_1\frac{m_e}{m_\tau}-\rho_2\right)m_\nu^{\text{1-loop}}
\end{align}
with the overall scale given by
\begin{equation}
m_\nu^{\text{1-loop}} \approx \frac{fy_\ell}{2(4\pi)^2}\frac{v_W\widetilde v}{M_h^2}\,m_\tau\,\ln\frac{M_h^2}{M_\phi^2}\;.
\end{equation}
Taking $\widetilde v \sim M_\phi \sim$ TeV and $M_h \sim 10$ TeV, we have $[v_W\widetilde v/(4\pi M_h)^2]m_\tau \sim 10^{4}\,\text{eV}$. The motivation behind this model is to take the Yukawa coupling $y_\ell \sim 1$, so we need
\begin{equation}\label{eq:loopcouplings}
|\rho_{1,2}f| \lesssim 10^{-4}
\end{equation}
to generate the scale $m_\nu \lesssim$ eV. The somewhat stringent requirement of Eq.~(\ref{eq:loopcouplings}) can be relaxed if we are willing to consider larger masses for the charged scalars, $h^\pm_{\al\beta}$, or smaller Yukawa couplings.
\\\\
Therefore, the $\Sigma(81)$ symmetry (along with the assumption that all new scales are roughly $\sim1-10$ TeV) suggests a relation among the off-diagonal entries of the neutrino mass matrix:
\begin{equation}\label{eq:prediction}
(m_\nu)_{\mu\tau} \approx \frac{m_\tau}{m_\mu}\,(m_\nu)_{e\mu}+\frac{m_\mu}{m_\tau}\,(m_\nu)_{e\tau}\;.
\end{equation}
The $\approx$ indicates that we have dropped small contributions due to the electron mass, which is justified provided that there is no large hierarchy between the couplings $\rho_1$ and $\rho_2$. The relation Eq.~(\ref{eq:prediction}) results from assuming that the only parameters contributing to $m_\nu^{\text{1-loop}}$ that break the flavor-independence predicted by $\Sigma(81)$ symmetry are the charged lepton masses $m_e \ll m_\mu \ll m_\tau$. The result Eq.~(\ref{eq:prediction}) is suggested by the marriage of the Private Higgs framework to the proposition that the fermion generations are interchangeable at high energy.
\\\\
In total we have a Majorana neutrino mass matrix\footnote{The most general Majorana mass matrix is symmetric, so we display explicitly only its upper triangle.}
\begin{equation}\label{eq:neutrinomassmatrix}
m_\nu \approx \frac{1}{m_0}\ml (v'_e)^2&0&0\\&(v'_\mu)^2&0\\&&(v'_\tau)^2 \mr+m_\nu^{\text{1-loop}}\ml 0&a&-b\\&0&r a\!-\!r^{-1}b\\&&0 \mr
\end{equation}
where $m_0 \equiv \sqrt2 y_N\widetilde v/y_\nu^2$, $a \approx r^{-1}\rho_1$, $b \approx \rho_2$, and $r \equiv m_\tau/m_\mu \approx 16.82$. Note that for the ``reasonable" ratio $\rho_1/\rho_2 \sim 1$ the mass matrix of Eq.~(\ref{eq:neutrinomassmatrix}) is not $\mu\tau$-symmetric, as expected on general grounds. 
\section{Comparison with oscillation data}\label{sec:data}
In a three-flavor oscillation framework with Majorana neutrinos, the three flavor eigenstates $\nu_e,\nu_\mu,\nu_\tau$ are linear combinations of the three mass eigenstates $\nu_1,\nu_2,\nu_3$, specified by a 3-by-3 unitary mixing matrix $V$, which has three physical mixing angles\footnote{In addition to the three angles in Eq.~(\ref{eq:expmixingangles}), the mixing matrix $V$ contains three physical complex phases and three unphysical complex phases. Of the three physical phases, only one is in principle observable in oscillations, the ``Dirac" phase angle $\delCP$. The remaining two physical phases are the ``Majorana" phases, which drop out of the oscillation probabilities.
}: the ``solar" angle $\ta_{12}$, the ``atmospheric" angle $\ta_{23}$, and the ``reactor" angle $\ta_{13}$, which lie in the following\footnote{The bounds on the reactor angle depend on a recent re-evaluation of the expected $\bar\nu_e$ flux; here we quote the least restrictive lower and upper bounds for $\ta_{13}$ in \cite{updatedreview}.} ranges  \cite{updatedreview}:
\begin{equation}\label{eq:expmixingangles}
31^\circ \leq \ta_{12} \leq 36^\circ\;,\;\; 36^\circ \leq \ta_{23} \leq 55^\circ\;,\;\; 7.2^\circ \leq \ta_{13} \leq 10^\circ\;.
\end{equation}
The oscillation phases also depend on the mass-squared differences $\msol \approx m_2^2-m_1^2$ and $\matm \approx m_3^2-m_2^2 \approx m_3^2-m_1^2$. Since the overall scale of the neutrino mass matrix is not known, the relevant observable to compare to data is the ratio of these $\Del m^2$s: 
\begin{equation}\label{eq:expR}
5.29 \leq \xi \equiv \sqrt{\frac{|\matm|}{\msol}} \leq 6.22
\end{equation}
for both possible orderings, $m_3 > m_2 > m_1$ (``normal") and $m_2 > m_1 > m_3$ (``inverted"). Using Eqs.~(\ref{eq:expmixingangles}) and~(\ref{eq:expR}), we now show that our model remains viable only for the ``normal" ordering, $m_3 > m_2 > m_1$.
\\\\
Define the quantity $\mathcal G \equiv (m_\nu)_{\mu\tau}-r(m_\nu)_{e\mu}-r^{-1}(m_\nu)_{e\tau}$, such that Eq.~(\ref{eq:prediction}) reads $\mathcal G \approx 0$, and solve this equation perturbatively in the small parameters in the problem: the deviation of the atmospheric angle from maximality, $-0.16\leq\ld \equiv \ta_{23}-\frac{\pi}{4}\leq+0.17$, and the small reactor angle, $0.13\leq \ta_{13} < 0.17$. Using $m_\nu = V^*\diag(m_1,m_2,m_3)V^\da$, with the standard angular parametrization for $V$, and expanding to leading order as\footnote{Note that the approximation $\mathcal G \approx \mathcal G_0$ is the condition of approximate $\mu\tau$ symmetry in the neutrino mass matrix.} $\mathcal G = \mathcal G_0+O(\ta_{13},\ld)$, we have
\begin{equation}\label{eq:G_0}
\mathcal G_0 = \half\left\{ \frac{\tilde m_1^*+\tilde m_2^*}{2}-m_3-F(r,\ta_{12})(\tilde m_2^*-\tilde m_1^*) \right\}\;,
\end{equation}
where
\begin{equation}\label{eq:F}
10.3 < F(r,\ta_{12}) \equiv \left[ \left(\frac{r+r^{-1}}{\sqrt2}\right)\sin(2\ta_{12})-\half \cos(2\ta_{12})\right] < 11.2\;.
\end{equation}
The tildes over $m_{1,2}$ in Eq.~(\ref{eq:G_0}) denote inclusion of the Majorana phases, and the range in Eq.~(\ref{eq:F}) result from applying Eq.~(\ref{eq:expmixingangles}). If the neutrino mass matrix is real, then $\tilde m_1 = \pm m_1$ and $\tilde m_2 = \pm m_2$ (signs uncorrelated).
\\\\
The sign of $\matm$ is presently unknown, so we should consider the cases $m_3 > m_2 > m_1$ (``normal" ordering) and $m_2>m_1>m_3$ (``inverted" ordering) separately. First consider the normal ordering, and take $m_3 \gg m_{1,2}$ in Eq.~(\ref{eq:G_0}). If we insist on $\mathcal G_0 \approx 0$ [in other words, satisfying Eq.~(\ref{eq:prediction}) with $\ta_{13} = 0$ and $\ta_{23} = \frac{\pi}{4}$], then we need $\tilde m_1\tilde m_2 < 0$ and $m_3 > m_{1,2}$. For example, if $\tilde m_1 = +m_1$ and $\tilde m_2 = -m_2$, then the condition $\mathcal G_0 \approx 0$ is satisfied with $m_3 \sim 10(m_1+m_2)$. 
\\\\
If instead $m_3 \ll m_{1,2}$, then $\mathcal G_0 \approx 0$ implies $\tilde m_1\tilde m_2 > 0$. Setting $\tilde m_1/m_1 = \tilde m_2/m_2 = s \equiv \pm 1$, we can solve explicitly for $m_1$ in terms of $F$ and the oscillation parameter $\xi$:
\begin{equation}\label{eq:m1inverted}
m_1 = \frac{\xi^2+F(F-1)+\fourth}{\sqrt{-(2F+1)\xi^2+2F^2-\half}}\sqrt{\frac{\msol}{2F-1}}
\end{equation}
where we have used $\matm = -|\matm|$ for $m_3 < m_{1,2}$. The sign $s$ has dropped out. Using the ranges in Eqs.~(\ref{eq:expR}) and~(\ref{eq:F}), we see that $-(2F+1)\xi^2+2F^2 < -300$ and so Eq.(\ref{eq:m1inverted}) has no solution. Nonzero values for $\ta_{13}$ and $\ld\equiv \ta_{23}-\frac{\pi}{4}$ are constrained by Eq.~(\ref{eq:expmixingangles}) to be small enough such that the model remains viable only for the ``normal" ordering, $m_3 > m_2 > m_1$.
\\\\
In the absence of a theoretical principle that determines the relative values of the diagonal entries in $m_\nu$, we proceed phenomenologically. If we take $v'_e \ll v'_\mu \sim v'_\tau$, then the mass matrix of Eq.~(\ref{eq:neutrinomassmatrix}) is similar to those of Class II in our earlier study \cite{Mee=0} of mass matrices with $(m_\nu)_{ee} = 0$. For example, 
\begin{equation}\label{eq:mnuexample}
m_\nu \propto \ml 0.01&0.25&-2\\&5&4.1\\&&5 \mr \implies \ta_{12} \approx 35.5^\circ\;,\;\; \ta_{23} \approx 43.5^\circ\;,\;\;\ta_{13} \approx 8.02^\circ\;,\;\;\xi \approx 5.8
\end{equation}
with $(m_1,m_2,m_3) \sim(1,\,1.6,\,7.2)$ in units of $m_1$. This is just meant to display a particular numerical example that is compatible with observation. 
\\\\
The conclusion to draw is simply that the Private Higgs model with $\sig$ symmetry can fit oscillation data with the ``reasonable" ratio of couplings $\rho_1/\rho_2 \sim O(1)$ [in the particular case of Eq.~(\ref{eq:mnuexample}) we have $\rho_1/\rho_2 \approx r/8 \approx 2.1$] and with $\matm > 0$. 
\\\\
In our model the diagonal entries of the Majorana neutrino mass matrix remain freely adjustable parameters, just like the charged fermion masses. We imagine that yet a further high-energy completion of the ``low-energy" PH model with $\sig$ symmetry, perhaps into the continuous Lie group $U(3)$, could shed light on the ``typical" values of these parameters.
\section{Lepton magnetic moments}\label{sec:g-2}
In our model, the dominant new contribution to the magnetic moment of charged leptons comes from neutral Higgs exchange.\footnote{The contribution to the muon magnetic moment from charged Higgs exchange is unobservably small. This can be seen from the diagram for $\mu \to 3e$ given in Fig.~\ref{fig:muto3e} with the neutral Higgs line removed, a photon line attached to the charged scalar propagator, and the $e^\da$ replaced by $\mu^\da$. Then the diagram has the correct chirality for a magnetic coupling $\sim F_{\mu\nu}\mu \s^{\mu\nu}\bar\mu+h.c.$ and involves a factor of $(m_\nu)_{\mu e}$ in the numerator.} In the Standard Model, the Higgs contributes to the anomalous magnetic moment $a_\al = \half(g_\al-2)$ of lepton $\al = e,\mu,\tau$ from the 1-loop diagram in Fig.~\ref{fig:magneticmoment}. For $m_H \gg m_\al$, this diagram yields the standard textbook result \cite{book}:
\begin{figure}[h]
\begin{center}
\fbox{
	\begin{minipage}{16 cm}
	\centering
	\includegraphics[width=80mm]{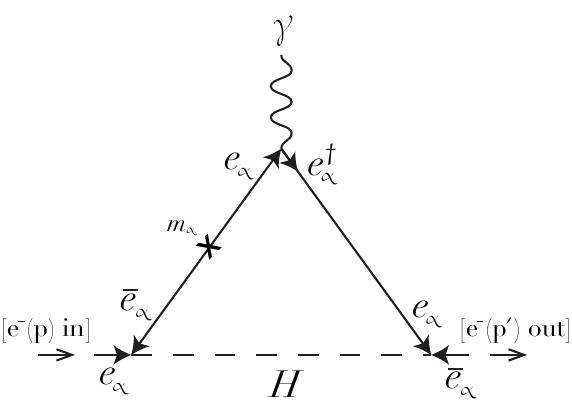}
	\caption{\small{One-loop contribution to the anomalous magnetic moment of lepton $\al = e,\mu,\tau$ in the SM due to Higgs exchange. We use two-component spinor notation for the diagram, in which each vertex is labeled by the fields that comprise the corresponding interaction in the Lagrangian, and the arrows denote spin: they point toward an undotted $(\half,0)$ index and away from a dotted $(0,\half)$ index in each vertex. The X in the $e_\al\bar e_\al$ line is a reminder that this propagator involves a power of lepton mass in the numerator and does not flip chirality, whereas the vectorlike interaction with the photon does flip chirality. There is also a second diagram in which the mass insertion is on the other leg of the triangle, and the electron-photon vertex is of the form $A_\mu \bar e_\al^\da \bar\s^\mu \bar e_\al$ (instead of $-A_\mu e_\al^\da \bar\s^\mu e_\al$). In our model, there are two leading contributions to the anomalous magnetic moment: an SM-like contribution from the mostly-$W$ Higgs, $H_W \approx H$, and a new contribution from the mostly-$\al$ Higgs, $H_\al$.}}
	\label{fig:magneticmoment}
	\end{minipage}
}
\end{center}
\end{figure}
\\
\begin{equation}\label{eq:magneticmomentSM}
a^{\text{SM}}_\al \approx \frac{y_\al^2}{(4\pi)^2}\frac{m_\al^2}{m_H^2}\,\ln\frac{m_H^2}{m_\al^2}\;.
\end{equation}
\\
In the SM, these contributions are negligible because of the small Yukawa couplings for charged leptons. In our model, the Yukawa couplings are independent of flavor and are also possibly $O(1)$. Moreover, when we diagonalize the scalar mass matrix, 
\begin{equation}
\ml H_W\\ H_e\\ H_\mu\\ H_\tau \mr = \U\ml H\\ H_1\\ H_2\\ H_3 \mr
\end{equation}
we find that each Private Higgs field is a linear combination of $H$, the SM Higgs, and the heavier mass eigenstates $H_i$. In particular, for the muon we have:
\begin{equation}
H_\mu \approx H_2+ \frac{y_\mu^{\text{SM}}}{y_\mu^{\text{PH}}}\left( 1+...\right) H+...
\end{equation}
where the first ellipsis denotes small (but possibly non-negligible) corrections to the SM Higgs-like coupling $\propto m_\mu/v$, and the second ellipsis stands for corrections due to quartic couplings of the form $\sim \phi_\al^\da \phi_\al \phi_\beta^\da \phi_\beta$, which are assumed to be negligibly small, as explained in the original Private Higgs proposal and as we will discuss below Eq.~(\ref{eq:Vsig}). The SM-like interaction reproduces Eq.~(\ref{eq:magneticmomentSM}), which can be ignored relative to the usual electroweak effects. 
\\\\
To evaluate the contribution from neutral $H_\mu$ exchange to $g-2$ of the muon, it is a sufficiently good approximation to simply think of the neutral Higgs boson that propagates in the loop for lepton $\al$ as being the Private Higgs boson $H_\al$, whose approximately flavor-independent mass $M_{\phi_\ell}$ is larger than the mass of the SM Higgs. Thus Eq.~(\ref{eq:magneticmomentSM}) is replaced by:
\begin{equation}\label{eq:magneticmoment}
a_\al \approx \frac{y_\ell^2}{(4\pi)^2}\frac{m_\al^2}{M_{\phi_\ell}^2}\,\ln\frac{M_{\phi_\ell}^2}{m_\al^2}\;.
\end{equation}
The contribution to the magnetic moment of the electron $a_e \sim 10^{-2}\times \left(\frac{10^{-4}\text{GeV}}{10^3\text{GeV}}\right)^2 \sim 10^{-16}$ is much larger than the Higgs contribution in the SM, but is still much smaller than the experimental uncertainty of $\sim 10^{-12}$. The contribution to the muon magnetic moment is a factor $m_\mu^2/m_e^2 \sim 10^6$ times larger, and is therefore in a range interesting for phenomenology:
\begin{equation}
a_\mu \sim (10^{-10}-10^{-9})\left(\frac{\text{TeV}}{M_{\phi_\ell}}\right)^2
\end{equation}
If we take seriously the reported 3$\s$ discrepancy from the SM prediction \cite{g-2anomaly}
\begin{equation}
\del a_\mu \equiv a^{\text{exp}}_\mu-a^{\text{SM}}_\mu = (4.3\pm 1.6)\times10^{-9}
\end{equation}
then our model can fit the deviation from the SM, provided that the Yukawa couplings are large and the Private Higgs mass is about a TeV. For example, $y_\ell = 1$ and $M_{\phi_\ell} = 530$ GeV, or\footnote{The value $y_\ell = 2$ is toward the upper limit for validity of perturbation theory but remains acceptable, which we argue as follows. If we approximate the beta function for the Yukawa coupling as $\mu\,dy_\ell/d\mu \approx a\,y_\ell^3/(4\pi)^2$, we can estimate the location $\Ld_\ell$ of the Landau pole, defined by $y^{-2}(\Ld_\ell) \equiv 0$. \cite{4gen} Using the value at $m_\tau$ as input, $y_\ell(m_\tau) = \sqrt2 m_\tau/v_\tau$, we find $\Ld_\ell = m_\tau\exp\left[ \frac{1}{a}\left(\frac{4\pi}{y_\ell(m_\tau)}\right)^2\right]$. The location of the pole is larger than $\sim10$ TeV provided that $y_\ell(m_\tau) \lesssim 4\pi\,a^{-1/2} \approx 8$, where we have used $a = \tfrac{5}{2}$. \cite{2HDM RG} This is of course a crude estimate, but the point is that relatively large values $y_\ell \sim 1$ can be considered.} $y_\ell = 2$ and $M_{\phi_\ell} = 1.1$ TeV, imply $a_\mu \approx 4.3\times10^{-9}$. For smaller values of $y_\ell$ and larger values of $M_{\phi_\ell}$, the contribution to the magnetic moment becomes smaller in magnitude than the electroweak contributions of the SM and can be dropped. Thus in general our model predicts that the anomalous magnetic moment of the muon is at most of order $\del a_\mu$ and at least as large as in the SM.
\section{Lepton flavor violation}\label{sec:LFV}
\begin{figure}[h]
\begin{center}
\fbox{
	\begin{minipage}{16 cm}
	\centering
	\includegraphics[width=120mm]{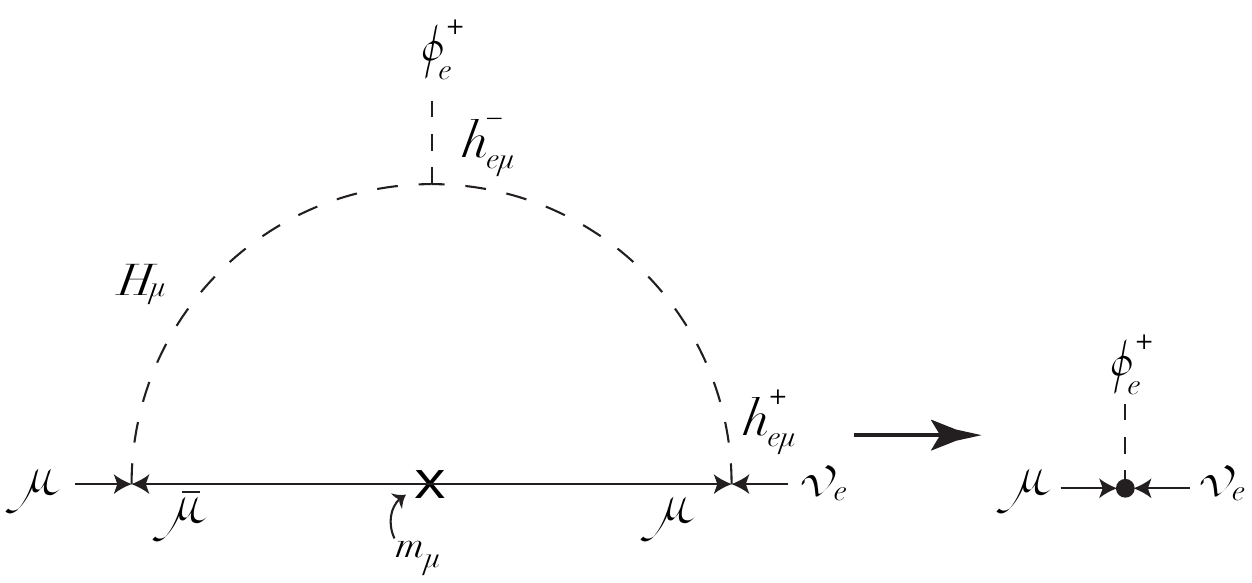}
	\caption{\small{The effective $\mu\nu_e\phi_e^+$ vertex at one loop. The arrow indicates taking the limit in which the scalars circulating in the loop are significantly more massive than the muon.}}
	\label{fig:y_eff}
	\end{minipage}
}
\end{center}
\end{figure}

\begin{figure}[h]
\begin{center}
\fbox{
	\begin{minipage}{16 cm}
	\centering
	\includegraphics[width=110mm]{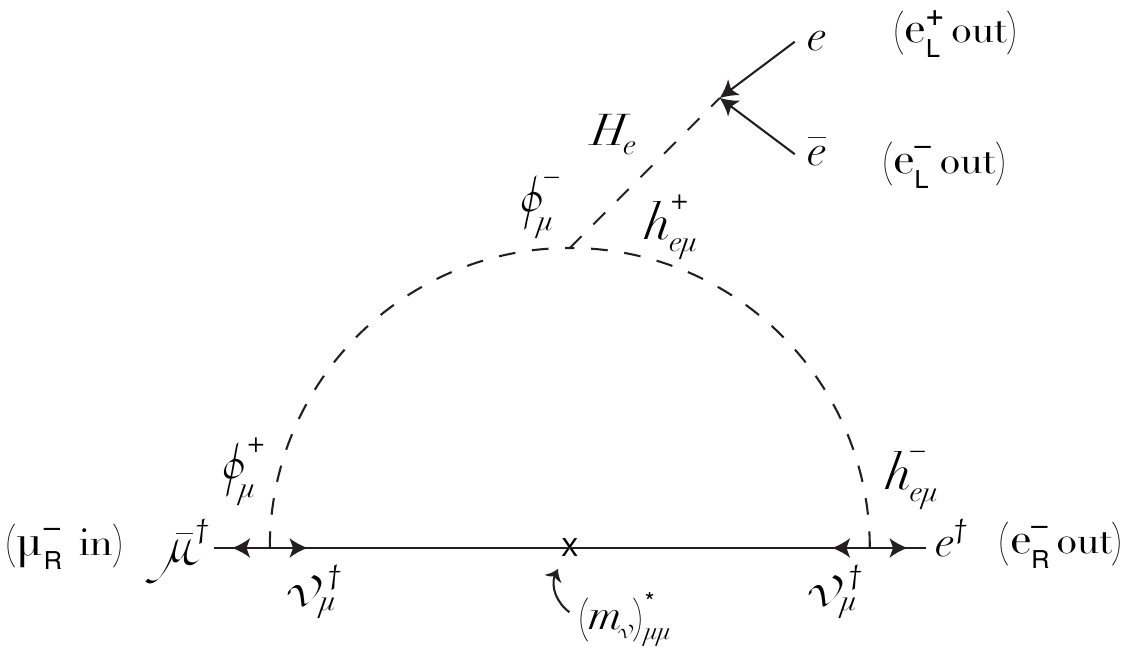}
	\caption{\small{One loop contribution to $\mu_R^- \to e_R^-e_L^+e_L^-$. The diagram for $\mu_R^- \to e_R^-\g$ is obtained by replacing the $H_e$ propagator with the VEV $\bra\phi_e^0\ket$ and attaching an external photon to the charged $h^\pm_{e\mu}$ or $\phi_\mu^\pm$ line. The X in the neutrino propagator denotes a power of neutrino mass in the numerator, which suppresses the rate significantly.}}
	\label{fig:muto3e}
	\end{minipage}
}
\end{center}
\end{figure}

Let us expand the lepton-Higgs doublets about their vacuum expectation values:
\begin{equation}
\phi_\al = \ml \tfrac{1}{\sqrt2}(v_\al+H_\al+i\chi_\al)\\ \phi_\al^- \mr\;.
\end{equation}
Here $H_\al$ and $\chi_\al$ are the physical CP-even and CP-odd parts, respectively, of the neutral component $\phi_\al^0$ of the lepton-Higgs doublets. Just as for the neutrino mass matrix, we will estimate the lepton flavor violating effects in the Higgs-flavor basis. The analysis of this section follows closely the phenomenology of the original PH model for leptons \cite{leptonPH}, with the notable exception that here we have $M_{\phi_e} \approx M_{\phi_\mu} \approx M_{\phi_\tau} \sim$ TeV instead of the original proposal $M_{\phi_e} \gg M_{\phi_\mu} \gg M_{\phi_\tau}$.
\\\\
The interaction $\hat Mh_{e\mu}^+\phi_e\cdot\phi_\mu$ from Eq.~(\ref{eq:hphi}) generates an effective vertex $\la = \tfrac{1}{\sqrt2}y_{\text{eff}}\,\mu\nu_e\phi_e^++h.c.$  at one loop through the diagram in Fig.~\ref{fig:y_eff}.
\\
\\
Including the exchange of the CP-odd state $\chi_\mu$, the dimensionless coefficient of this effective vertex is 
\begin{equation}
y_{\text{eff}} \approx \frac{fy_\ell}{(4\pi)^2}\frac{\hat Mm_\mu}{M_h^2}\,\ln\frac{M_h^2}{M_{\phi_\ell}^2}\;.
\end{equation}
Integrating out $\phi_e^\pm$ leads to the effective interaction $\la = 4\sqrt2 G_F^{(\mu\to\nu_e \bar\nu_ee)}(\mu\nu_e)(\nu_e\ebar)+h.c.$ with a 4-fermi constant
\begin{equation}
G_F^{(\mu\to\nu_e \bar\nu_ee)} \approx \frac{fy_\ell^2}{128\pi^2}\frac{\hat Mm_\mu}{M_{\phi_\ell}^2M_h^2}\ln\frac{M_h^2}{M_{\phi_\ell}^2}\;.
\end{equation}
As denoted, this leads to a flavor changing charged current (FCCC) contribution to muon decay. Compared with the SM muon decay Lagrangian $\la = 4\sqrt2 G_F(\mu\nu_e)(\nu_\mu^\da e^\da)+h.c.$ with $G_F = 1.166\times 10\,\text{TeV}^{-2}$, the new contribution is suppressed by the ratio
\begin{equation}
\frac{G_F^{(\mu\to\nu_e \bar\nu_ee)}}{G_F} \sim \frac{f\hat M\,\text{TeV}}{10\,M_h^2}\;.
\end{equation}
where we have taken $y_\ell \sim 1$, $M_{\phi_\ell} \sim$ TeV, and $\ln\frac{M_h^2}{M_{\phi_\ell}^2} \sim 1$. For example, taking $\hat M \sim$ TeV and $M_h \sim 10$ TeV, requiring that the above ratio be less than $10^{-3}$ implies only the rather weak constraint that $f < 1$, which is readily satisfied in our model.
\\\\
Another non-standard contribution to FCCC-mediated muon decay comes from integrating out the charged boson $h^+_{e\mu}$. Dropping small mixing terms with $\phi_e^+$ and $\phi_\mu^+$, the interactions of Eq.~(\ref{eq:hll}) result in the effective Lagrangian
\begin{equation}
\la_{\text{eff}} \approx  \frac{f^2}{M_h^2}(\nu_e\mu)(\nu_\mu^\da e^\da)+h.c.
\end{equation}
Therefore the Fermi constant measured from muon decay gets an extra contribution of order $\sim \frac{f^2}{M_h^2}/G_F$ compared to the value of $G_F$ extracted from hadronic weak decays. Requiring the discrepancy between the two to be less than $10^{-3}$ results in the constraint \cite{mitsuda}
\begin{equation}
M_h > f\times(10^3G_F^{-1})^{1/2} \sim 10f\,\text{TeV}\;.
\end{equation}
If we take $M_h \sim 10$ TeV, we can satisfy this equation even for a coupling as large as $f \sim 1$.
\\\\
Potentially dangerous flavor changing neutral currents (FCNC) arise from diagrams similar to the above, but with a neutrino running in the loop instead of a charged lepton. These amplitudes are therefore suppressed by a neutrino mass in the numerator and thereby result in negligible branching fractions.
\\\\
For example, the polarized muon decay $\mu_R^- \to e_R^- e_L^+e_L^-$ arises from the one loop diagram in Fig~\ref{fig:muto3e}.
\\
\\
This diagram, along with the diagram for $\chi_e$ exchange, results in the effective Lagrangian
\begin{equation}
\la_{\text{eff}}^{\text{1-loop}} = 4\sqrt2G_F^{(\mu\to3e)}(\mubar^\da e^\da)(e\ebar)+h.c.
\end{equation}
where the effective Fermi coupling is
\begin{equation}
G_F^{(\mu\to3e)} \approx \frac{fy_\ell^2}{128\pi^2}\,\frac{\hat M(m_\nu)^*_{\mu\mu}}{M_{\phi_\ell}^2M_h^2}\,\ln\frac{M_h^2}{M_{\phi_\ell}^2}\;.
\end{equation}
This is identical in form to the FCCC coupling discussed previously, suppressed by the ratio
\begin{equation}
\frac{G_F^{(\mu\to3e)}}{G_F^{(\mu\to\nu_e\bar\nu_ee)}} \sim \frac{m_\nu}{m_\mu} \sim 10^{-10}-10^{-8}\;.
\end{equation}
With $G_F^{(\mu\to\nu_e\bar\nu_ee)}/G_F \lesssim 10^{-3}$, the $\mu \to 3e$ decay is suppressed by at least a factor $\sim 10^{-11}$ at the level of the amplitude, resulting in a branching fraction less than $10^{-22}$, which is much smaller than the existing upper bound of $\sim 10^{-12}$ \cite{B(mutoeee)}. Other FCNC processes such as $\mu^- \to e^-\g$ and $Z \to \mu^-\tau^+$ are similarly suppressed by a factor of $m_\nu$ and are therefore unobservably small \cite{mutoegammasuppressedbymnu}.
\section{Quark masses}\label{sec:quarks}
The basic structure of this model can be carried over immediately to the quark sector. Introducing Private Higgs fields $\phi_D \sim (2,-\half)$ and $\phi_U \sim (2,+\half)$ for the quarks, we can identify the quark sector with the lepton sector through the correspondence:
\begin{equation}
\ml \ell \\ \overline e\\ N\\ \phi_\ell \\ \phi_\nu \mr \lrarrow \ml q\\ \overline d\\ \overline u\\ \phi_D\\ \phi_U \mr\;.
\end{equation}
Just as $\phi_\ell = (\phi_e,\phi_\mu,\phi_\tau)$ and $\phi_\nu = (\phi_{\nu_e},\phi_{\nu_\mu},\phi_{\nu_\tau})$ are flavor triplets, here we have $\phi_D = (\phi_d,\phi_s,\phi_b)$ and $\phi_U = (\phi_u,\phi_c,\phi_t)$. To prevent the lepton-Higgs fields from coupling to the quarks and the quark-Higgs fields from coupling to leptons, the triplets in the quark sector should not be the defining triplet 3 but rather one of the other three complex triplet representations, $3'$, $\hat3$, or $\tilde3$ [see Table~\ref{table:generators on 3} and Eq.~(\ref{eq:triplet notation})]. Since we have already assigned the spin-0 charged bosons $h^-$ to the $3'$, we have the choice of either $\hat 3$ or $\tilde 3$ for the quark triplets. 
\\\\
Interestingly, the product $\hat 3\x \hat 3\x \hat3$ contains three different invariants of $\sig$ [see Eq.~(\ref{eq:3''x3''=...})], so if the quarks and quark-Higgs were assigned to the $\hat 3$ there would be three Yukawa couplings for the down-type quarks and three for the up-type quarks. Instead, since $\tilde 3\x \tilde3 = \tilde 3^*\+3\+3$ [see Eq.~(\ref{eq:3'''x3'''=...})], the product $\tilde 3\x \tilde3\x \tilde3$ contains only one invariant, just like the product $3\x3\x3$ for leptons. Therefore, we assign the quarks and quark-Higgs fields to the $\tilde 3$ representation.
\\\\
The quark Yukawa couplings are
\begin{equation}
\la_{\text{Yuk}}^q = y_D(q_1\!\cdot\!\phi_d\, \overline d+q_2\!\cdot\!\phi_s\,\overline s+q_3\!\cdot\!\phi_b\,\overline b)-y_U(q_1\!\cdot\!\phi_u\,\overline u+q_2\!\cdot\!\phi_c\,\overline c+q_3\!\cdot\!\phi_t\,\overline t)+h.c.
\end{equation}
The quark masses are therefore determined by Private Higgs VEVs\footnote{Note that since $m_t \approx 173$ GeV is larger than $m_W$, the $W$ boson gets contributions from both the $t$-Higgs and the $W$-Higgs, so that Eq.~(\ref{eq:mW}) gets modified to
\begin{equation}
m_W \approx \half g(v_W^2+v_t^2)^{1/2}\left[1+O\!\left(\frac{v_{\text{other}}^2}{v_W^2+v_t^2}\right)\right]\;.
\end{equation}
This just means that $v_{\text{SM}} \approx (v_W^2+v_t^2)^{1/2} \approx 246$ GeV gets roughly half its contribution from each of $v_W$ and $v_t$ rather than purely from $v_W$. In contrast to the other Private Higgs fields, the top-Higgs mixes substantially with the $W$-Higgs, which should be interesting for LHC phenomenology.} and two generation-independent couplings, one for the down-type quarks and one for the up-type quarks:
\begin{equation}
m_{d,s,b} = y_D\bra\phi_{d,s,b}^0\ket\qquad\text{ and }\qquad m_{u,c,t} = y_U\bra\phi_{u,c,t}^0\ket\;.
\end{equation}
The quark mass matrices for the up and down type quarks are simultaneously diagonal, and so the CKM matrix is the identity to leading order. We can then generate nonzero mixing angles at one loop by introducing the appropriate colored fields.\footnote{The idea of generating hierarchical quark masses and mixing angles at one, two, and three loops was explored by Babu and Mohapatra in the context of an $S_3$ model \cite{babu}. An approach along the lines of our model would attempt to generate a realistic CKM matrix at one loop, in analogy with neutrino mixing.} Just as the $h^-_{\al\beta}$ transform as the components of a triplet $3'$ distinct from the $3$, these new colored fields will transform as the components of a triplet. In view of $\tilde 3\x \tilde3 = \tilde 3^*\+3\+3$, the colored fields will have to transform as the defining $3$, the same representation assigned to the leptons.\footnote{Color $SU(3)$ invariance will prohibit tree level interactions, but it is conceivable that loop level interactions may still have interesting implications for neutrino mixing \cite{classification}. } Introducing other fields that transform as the remaining triplet $\hat 3$ may have interesting phenomenological consequences [see Eqs.~(\ref{eq:3x3''=...}),~(\ref{eq:3x(3'')*=...}),~(\ref{eq:3''x3'''=...}) and~(\ref{eq:3''x(3''')*=...})].
\\\\
The immediate phenomenological challenge to the possibility of adapting this model for the quark sector has to do with the flavor changing requirements in the quark sector, which are significantly more stringent than those for the leptons. Since our emphasis here is the lepton sector, we will postpone the construction of a realistic CKM matrix to future work, and so a full discussion of flavor physics in the quark sector is beyond the scope of this paper. 
\\\\
The purpose here is to present an alternative approach to the quark flavor puzzle: in the SM, the quark mass matrices are in principle arbitrary, and so the question is why they should be approximately diagonal in the same basis. In the PH model, the quark mass matrices are engineered to be diagonal, and the question arises as to why $V_{\text{CKM}}$ should deviate from the unit matrix at all.
\section{Scalar potential}\label{sec:potential}
Let us now discuss the scalar potential for our model. The most general scalar potential for the fields $\phi_W,\phi_\ell,\phi_\nu$, and $S$ invariant under $SU(2)\times U(1)\times\sig$ is\footnote{Note the presence of the flavor-diagonal $S_\al^3$ terms for the SM-singlet scalars. This is in contrast to the original Private Higgs model, for which such terms were forbidden by $\zb_2$ parities. A similar approach was also used by E. Ma in the context of a different model \cite{Z3DMand2loopmnu}.}:
\begin{align}\label{eq:Vsig}
V_{\sig} &= M_{\phi_W}^2\phi_W^\da\phi_W+M_{\phi_\ell}^2\sal \phi_\al^\da\phi_\al+M_{\phi_\nu}^2\sal \phi_{\nu_\al}^\da\phi_{\nu_\al}+M_S^2\sal S_\al^\da S_\al \nonumber\\
&+\sqrt2\left(-\mu\sal S_\al\phi_\al^\da\phi_W-\mu'\sal S_\al^\da\phi_{\nu_\al}\!\!\!\cdot\!\phi_W+\tfrac{1}{3}\widetilde\mu\sal S_\al^3+h.c.\right) \nonumber\\
&+\ld_W(\phi_W^\da\phi_W)^2+\ld_S(\sum_\al S_\al^\da S_\al)^2+\ld_S'(S_e^\da S_eS_\mu^\da S_\mu+\text{cyclic})+V_{\sig}^{\text{quartic}}
\end{align}
where $V^{\text{quartic}}_{\sig}$ contains all allowed quartic interactions other than those displayed explicitly in Eq.~(\ref{eq:Vsig}). We assume that all of the couplings in $V^{\text{quartic}}_{\sig}$ are of the correct sign so as to stabilize the potential for large values of the fields, but that they are otherwise negligible for perturbative dynamics. In other words, just as in the original PH model \cite{PH, leptonPH}, we assume that it is a good approximation\footnote{For Yukawa couplings of order 1, one might worry that negligible quartic couplings at $\mu\sim$ TeV could blow up at low energy. If $y_\ell \sim 1$, the leading order contribution to the renormalization of $\ld_\al(\phi_\al^\da\phi_\al)^2$ comes from a box diagram with four external $\al$-Higgs lines and the lepton $\al$ running in the loop \cite{higgsRG}. This diagram is $O(y_\ell^4)$, so for large Yukawa couplings we approximate the beta function for $\ld_\al$ as $(4\pi)^2 \mu\,d\ld_\al/d\mu \approx -b\,y_\ell^4$ with $b > 0$. (This is just a crude approximation; $\ld \phi^4$ theory coupled to fermions is not asymptotically free.~\cite{zeeRG}) Approximating $(4\pi)^2\mu\,dy_\ell/d\mu \approx a\,y_\ell^3$ as before, we find $\ld_\al(\mu)-\ld_\al(\mu_0) = -\frac{b}{2a}[y_\ell^2(\mu)-y_\ell^2(\mu_0)]$ $= -\frac{b}{2a}y_\ell^2(\mu_0)\left\{ \left[ 1-a[y_\ell(\mu_0)/4\pi]^2\ln(\mu/\mu_0)\right]^{-1}-1 \right\}$. For example, taking $\al = e$ with $a = \tfrac{5}{2}$ and $b = 4$ \cite{2HDM RG} and running from $\mu_0 =$ TeV to $\mu = m_e$, we have $\ld_e(m_e)-\ld_e(\text{TeV}) = O(10^{-5})$ for $y_\ell(\text{TeV}) = 0.1$, $\ld_e(m_e)-\ld_e(\text{TeV}) = 0.15$ for $y_\ell(\text{TeV}) = 1$, and $\ld_e(m_e)-\ld_e(\text{TeV}) = 1.5$ for $y_\ell(\text{TeV}) = 2$. In any case, as long as the theory remains perturbative, a term $\fourth \ld_ev_e^4$ in Eq.~(\ref{eq:Vsig}) would have no qualitative effect on the dynamics of the theory, partially because $v_e \ll v_W$ and partially because the interaction is diagonal in flavor. Flavor off-diagonal effective interactions $(\phi_W^\da\phi_W)(\phi_\al^\da\phi_\al)$ and $(\phi_\al^\da\phi_\al)(\phi_\beta^\da\phi_\beta)$ do not receive contributions from Yukawa couplings at leading order, and therefore can remain small for all scales in the model.} to ignore the terms in $V^{\text{quartic}}_{\sig}$.
\\\\
The potential $V_{\sig}$ is minimized when all VEVs are independent of flavor, and so the charged leptons would be degenerate in mass. This is of course phenomenologically unacceptable. To fix this problem, we would like to forbid the bare cubic terms $S_\al\phi_\al^\da\phi_W$ and $S_\al^\da\phi_{\nu_\al}\!\!\!\cdot\!\phi_W$, and then re-introduce these terms with parametrically smaller, flavor-dependent couplings $\mu_\al$ and $\mu'_\al$. 
\\\\
To do this, we impose an auxiliary flavor-independent $\zb_3$ symmetry, denoting its generator by $\W \equiv e^{\,i2\pi/3}$. Under this $\zb_3^\W$, the Higgs doublets $\phi_\ell$ and $\phi_\nu$ transform as $\W$, while the fields $S$, $\ebar$, and $N$ transform as $\W^*$. The lepton doublets $\ell$ and the spin-0 charged bosons $h^-$ do not transform under $\zb_3^\W$. Note that this $\zb_3^\W$ also forbids the interaction $h^+_{\al\beta}\phi_\al\!\cdot\!\phi_\beta$.
\\\\
The sequence of flavor symmetry breaking should be
\begin{equation}\label{eq:symmetrybreakingsequence}
\sig\times\zb_3^\W\;\; \begin{matrix}\M\\ \to\\ \phantom{1} \end{matrix}\;\; (\zb_3^e\times\zb_3^\mu\times\zb_3^\tau)\times \zb_3^\W \;\; \begin{matrix}\M'\\ \to\\ \phantom{1} \end{matrix}\;\;  \zb_3^e\times\zb_3^\mu\times\zb_3^\tau \;\; \begin{matrix}\M''\\ \to\\ \phantom{1} \end{matrix}\;\;  \emptyset\;.
\end{equation}
for some scales $\M \sim \M' > \M'' \gtrsim 10^2$ GeV. At the scale $\M$, the full $\sig = (\zb_3^e\times\zb_3^\mu\times\zb_3^\tau)\rx\zb_3^C$ symmetry is broken to its abelian discrete subgroup $\zb_3^e\times\zb_3^\mu\times\zb_3^\tau$, but the unbroken $\zb_3^\W$ symmetry still forbids the interactions $S_\al\phi_\al^\da\phi_W$ and $S_\al^\da\phi_{\nu_\al}\!\!\!\cdot\!\phi_W$. Then at the scale $\M'$, the $\zb_3^\W$ is broken, generating small (relative to $\M')$ coefficients $\mu_\al$ and $\mu'_\al$ for $S_\al\phi_\al^\da\phi_W$ and $S_\al^\da\phi_{\nu_\al}\!\!\!\cdot\!\phi_W$, respectively. Since $\sig$ is broken, these coefficients can in principle depend strongly on flavor, as indicated by the label $\al = e,\mu,\tau$. It is possible for the two scales to coincide, as we will show in a particular example. Finally, the scale $\M''$ corresponds to the largest VEV that breaks any of the flavor-dependent $\zb_3^\al$, namely the largest $\bra S_\al\ket$.
\\\\
Therefore, the potential we use in our model is
\begin{equation}\label{eq:potential}
V = V_{\sig\times\zb_3^\W}\;+\;V_{\text{soft}}+...
\end{equation}
where
\begin{align}\label{eq:invariantpotential}
&V_{\sig\times\zb_3^\W} = M_{\phi_W}^2\phi_W^\da\phi_W+M_{\phi_\ell}^2\sal \phi_\al^\da\phi_\al+M_{\phi_\nu}^2\sal \phi_{\nu_\al}^\da\phi_{\nu_\al}+M_S^2\sal S_\al^\da S_\al \\
&\phantom{V_{\sig\times\zb_3^\W}}+\sqrt2\left(\tfrac{1}{3}\widetilde\mu\sal S_\al^3+h.c.\right)+\ld_W(\phi_W^\da\phi_W)^2+\ld_S(\sum_\al S_\al^\da S_\al)^2+\ld_S'(S_e^\da S_eS_\mu^\da S_\mu+\text{cyclic})\nonumber
\end{align}
and
\begin{equation}\label{eq:softpotential}
V_{\text{soft}} = -\sqrt2\sal\left(\phantom{\frac{1}{1}}\!\!\!\! \mu_\al S_\al\phi_\al^\da\phi_W+\mu'_\al S_\al^\da \phi_{\nu_\al}\!\!\!\cdot\!\phi_W+h.c.\right)\;.
\end{equation}
The ellipsis in Eq.~(\ref{eq:potential}) denotes quartic Higgs-Higgs, scalar-scalar, and Higgs-scalar interactions, which we assume to be negligible, and small flavor-dependent corrections to the bare parameters contained in $V_{\sig\times\zb_3^\W}$. We imagine that flavor-dependent corrections are of the order $\mu_\al^{\text{max}} \sim (\mu'_\al)^{\text{max}} \sim \e\M'$, where $\e$ is a small number parametrizing the effect of either a small coupling or a loop factor $1/(4\pi)^2 \sim 10^{-2}$. Thus taking $\M' \sim$ TeV, we take the largest of the dimensionful flavor-asymmetric parameters to be $\mu_\al^{\text{max}} \sim (\mu'_\al)^{\text{max}} \sim 10$ GeV $\sim 10\,m_\tau$.
\\\\
Before discussing a particular realization of this idea, let us first treat Eqs.~(\ref{eq:potential}),~(\ref{eq:invariantpotential}), and~(\ref{eq:softpotential}) as an effective field theory and work out its consequences for the boson masses.
\\\\
We choose the basis of VEVs such that $\bra\phi_W^0\ket \equiv \tfrac{1}{\sqrt2}v_W$ is real. We will also fix unitary gauge, in which the Goldstone bosons $G^0 \equiv v_W\chi_W+\sum_\al(v_\al\chi_\al+v'_\al\chi_{\nu_\al})$ and $G^\pm \equiv v_W\phi_W^\pm+\sum_\al (v_\al\phi_\al^\pm+v'_\al\phi_{\nu_\al}^\pm)$ are set to zero. Since $v_W$ is much larger than $v_\al$ and $v'_\al$, in practice fixing unitary gauge amounts to setting $\chi_W$ and $\phi_W^\pm$ approximately equal to zero, up to corrections at most of order $v_\tau/v_W \sim m_\tau/m_W \sim 10^{-2}$:
\begin{equation}
\phi_W \approx \ml \tfrac{1}{\sqrt2}(v_W+H_W)\\ 0 \mr\qquad(\text{unitary gauge})\;.
\end{equation}
As far as the $SU(2)\times U(1)$ gauge interactions are concerned, the real field $H_W$ is the SM Higgs up to small corrections. Expanding the other Higgs doublets about their VEVs, we write
\begin{equation}
\phi_\al = \ml \tfrac{1}{\sqrt2}(v_\al\,e^{\,i\ta_\al}+H_\al+i\chi_\al)\\ \phi_\al^- \mr\;,\qquad \phi_{\nu_\al} = \ml \phi_{\nu_\al}^+\\ \tfrac{1}{\sqrt2}(v'_\al\,e^{\,i\ta'_\al}+H_{\nu_\al}+i\chi_{\nu_\al}) \mr\;.
\end{equation}
For simplicity, we assume that the VEVs are real: $\ta_\al = \ta'_\al = 0$. We also have
\begin{equation}
S_\al = \tfrac{1}{\sqrt2}\widetilde v_\al\,e^{\,i\widetilde\ta_\al}+s_\al
\end{equation}
where the $s_\al$ are physical complex SM-singlet spin-0 bosons. Again for simplicity we set $\widetilde\ta_\al = 0$. We will leave the study of CP-violating phenomenology for future work.
\\\\
The mostly-$H_W$ state serves as the SM-like Higgs boson, and has a squared mass $\sim \ld_Wv_W^2$. The mostly-PH states have masses determined by minimizing the potential to be
\begin{equation}\label{eq:privatehiggsmass}
M_{\phi_\ell}^2 = \frac{\mu_\al \widetilde v_\al v_W}{v_\al}\qquad\text{ and }\qquad M_{\phi_\nu}^2 = \frac{\mu'_\al \widetilde v_\al v_W}{v'_\al}\;.
\end{equation}
To the extent to which the bare parameters $M_S^2$ and $\widetilde\mu$ are independent of flavor, the scalar VEVs are independent of flavor: 
\begin{equation}\label{eq:scalar vevs equal}
\widetilde v_e \approx \widetilde v_\mu \approx \widetilde v_\tau \equiv \widetilde v.
\end{equation}
To the same extent to which the bare parameters $M_{\phi_\ell}^2$ and $M_{\phi_\nu}^2$ are independent of flavor [see, e.g., Eq.~(\ref{eq:flavor dependence of PH mass})], we are forced to determine the Higgs VEVs $v_\al$ and $v'_\al$ by appropriately tuning the soft parameters $\mu_\al$ and $\mu'_\al$:
\begin{equation}\label{eq:higgs vevs hierarchical}
\mu_e:\mu_\mu:\mu_\tau = v_e:v_\mu : v_\tau = m_e:m_\mu :m_\tau\;.
\end{equation}
Similarly, the ratios $\mu'_e:\mu'_\mu :\mu'_\tau = v'_e:v'_\mu:v'_\tau$ determine the diagonal entries in the Majorana mass matrix for the light neutrinos. In contrast to the situation for charged lepton masses, we do not know the required hierarchy of these entries. We could have $v'_e \sim v'_\mu \sim v'_\tau$, or one of the entries much smaller than the rest, for example $v'_e \ll v'_\mu \sim v'_\tau$. This is, of course, an accommodation of the empirical data rather than a prediction of it. The theoretical distinction from previous work (e.g. the original Private Higgs model) is that the charged lepton masses are determined by a hierarchy in ``soft" parameters $\mu_\al$ rather than by a hierarchy in PH masses.
\\\\
Taking $\mu_\tau/v_\tau \sim 10$, $\tilde v \sim$ TeV and $v_W \sim 10^{2}$ GeV, we find
\begin{equation}
M_{\phi_\ell}^2 \sim (\text{TeV})^2\;.
\end{equation}
This is consistent with our prediction for the anomalous magnetic moment of the muon.
\\\\
For the sake of not having scales higher than $\sim 10$ TeV, then $\widetilde v\,v_W \sim 10^{-1}\text{ TeV}^2$ implies that we should take $\mu'_\al/v'_\al$ no larger than $\sim 10^3$. It is consistent to take $\mu'_\tau \sim \mu_\tau \sim 10$ GeV and $v'_\tau \sim 10^{-3}\mu'_\tau \sim 10$ MeV. We can then take $v_\mu' \sim (1-10)$ MeV and $v'_e \ll$ MeV, in a spirit similar to $v_e \ll v_\mu \sim 10^{-1} v_\tau$ for the charged leptons.
\section{An example of soft symmetry breaking}\label{sec:soft}
Now we discuss one way to obtain the potential of Eq.~(\ref{eq:potential}). The reader is encouraged to find alternative high-energy completions. 
\\\\
Let $\s_0$, $\s_1$, and $\s_2$ be SM-singlet complex scalar fields that transform with the phase $\W$ under the auxiliary symmetry $\zb_3^\W$. Let these fields also transform as $\s_0 \sim 1$, $\s_1 \sim 1'$, and $\s_2 \sim (1')^*$ under $\sig$. We introduce the usual negative mass-squared instability for each of the $\s_i$ so that they obtain nonzero VEVs. 
\\\\
The fields $\s_1$ and $\s_2$ are charged under $\sig$ but are invariant under the abelian subgroup $\zb_3^e\times\zb_3^\mu\times\zb_3^\tau$ (see Table~\ref{table:generators on 1} in the appendix), and all $\s_i$ are charged under $\zb_3^\W$. Therefore, when the $\s_i$ obtain VEVs the flavor symmetry breaks as $\sig\times\zb_3^\W\to\zb_3^e\times\zb_3^\mu\times\zb_3^\tau$. In the notation of Eq.~(\ref{eq:symmetrybreakingsequence}), we have $\M = \M' = \max(\bra\s_0\ket,\bra\s_1\ket,\bra\s_2\ket)$. 
\\\\
Since $3\x3^* = 1\+1'\+(1')^*\+...$ in $\sig$ [see Eq.~(\ref{eq:3 x 3* decomposition})], the fields $\s_i$ allow the $[SU(2)\times U(1)\times \sig\times\zb_3^\W]$-invariant interactions $\ld_0\s_0^\da (S\phi^\da)_1+\ld_1\s_1^\da (S\phi^\da)_{1'}+\ld_2\s_2^\da (S\phi^\da)_{1''}]\phi_W$ = $(\ld_0\s_0^\da \!+\!\ld_1\s_1^\da \!+\!\ld_2\s_2^\da )S_e\phi_e^\da\phi_W\!+\!(\ld_0\s_0^\da \!+\!\w\ld_1\s_1^\da \!+\!\w^*\ld_2\s_2^\da )S_\mu\phi_\mu^\da\phi_W\!+\!(\ld_0\s_0^\da \!+\!\w^*\ld_1\s_1^\da \!+\!\w\ld_2\s_2^\da )S_\tau\phi_\tau^\da\phi_W$.
\\\\
When the $\s_i$ pick VEVs, we obtain the soft potential of Eq.~(\ref{eq:softpotential}) with dimensionful couplings
\begin{align}
&\mu_e = \tfrac{1}{\sqrt2}(\ld_0\bra\s_0\ket^*+\ld_1\bra\s_1\ket^*+\ld_2\bra\s_2\ket^*)\;, \nonumber\\
&\mu_\mu = \tfrac{1}{\sqrt2}(\ld_0\bra\s_0\ket^*+\w\ld_1\bra\s_1\ket^*+\w^*\ld_2\bra\s_2\ket^*)\;, \nonumber\\
&\mu_\tau = \tfrac{1}{\sqrt2}(\ld_0\bra\s_0\ket^*+\w^*\ld_1\bra\s_1\ket^*+\w\ld_2\bra\s_2\ket^*)\;.
\end{align}
\\
We would like these couplings to be at most a few orders of magnitude smaller than the weak scale. Even if $\bra\s_i\ket \sim (1-10)$ TeV $\gtrsim\bra S_\al\ket \sim$ TeV, we can have $\mu_\al\lesssim 10$ GeV if the couplings\footnote{In terms of scales, this is analogous to the soft breaking of chiral symmetry in QCD. The SM Higgs VEV $v \sim 10^2$ GeV is much larger than the chiral symmetry breaking scale $\xi \sim 10^2$ MeV, while the light quark masses $m_q = \tfrac{1}{\sqrt2}y_qv \sim (1-10^2)\text{MeV}$, which break $SU(3)_L\times SU(3)_R$ softly to the diagonal subgroup $SU(3)_V$, are much smaller than $\xi$ despite being proportional to $v$.} $\ld_i$ are of order $\sim 10^{-3}-10^{-2}$.
\\\\
Note that if the $\ld_i$ and $\bra\s_i\ket$ are all real, then $\mu_\mu = \mu_\tau$ but $\mu_{\mu,\tau} \neq \mu_e$. This can be viewed as a motivation for taking the second and third generation $\mu_\al$ (and $\mu'_\al$) as ``comparable," while treating $\mu_e$ and $\mu'_e$ as somehow ``special." This is an attractive scenario to explain the neutrino Higgs VEVs $v'_\mu \approx v'_\tau$, but the charged lepton relation $v_\mu/v_\tau = m_\mu/m_\tau \approx 5.9\%$ requires some fine tuning, as in many flavor models based on non-abelian groups.
\\\\
Note also that the field $\s_0$ allows the dimension-4 interaction
\begin{equation}
\ld_h\s_0(h^+_{\mu\tau}\phi_\mu\!\cdot\!\phi_\tau+h^+_{\tau e}\phi_\tau\!\cdot\!\phi_e+h^+_{e\mu}\phi_e\!\cdot\!\phi_\mu)\;.
\end{equation}
The VEV of $\s_0$ generates the dimension-3 interaction of Eq.~(\ref{eq:hphiphi}) with coefficient $\hat M = \ld_h\bra\s_0\ket$. The scale $\bra\s_0\ket \sim (1-10)$ TeV times a coupling $\ld_h \sim 10^{-1}-1$ results in $\hat M \sim$ TeV, which is the case studied in the previous sections.
\\\\
Furthermore, the fields $\s_0$, $\s_1$, and $\s_2$ result in small flavor-dependent corrections to the PH masses. In particular, the operator $(\phi_\ell^\da\phi_\ell)_1$ couples to $|\s_0|^2$, $|\s_1|^2$, and $|\s_2|^2$, the operator $(\phi_\ell^\da\phi_\ell)_{1'}$ couples to $\s_0\s_1^\da$ and $\s_0^\da\s_2$, and the operator $(\phi_\ell^\da\phi_\ell)_{(1')^*}$ couples to $\s_0^\da\s_1$ and $\s_0\s_2^\da$. Explicitly, defining $A \equiv (a_0|\s_0|^2\!+\!a_1|\s_1|^2\!+\!a_2|\s_2|^2)$, $B \equiv  (b_1\s_0\s_1^\da\!+\!b_2\s_0^\da\s_2)$, and $C \equiv (c_1\s_0^\da\s_1\!+\!c_2\s_0\s_2^\da)$, we have:
\begin{align}
&A(\phi_\ell^\da\phi_\ell)_{1}+\half\left[B(\phi_\ell^\da\phi_\ell)_{1'}\!+\!C(\phi_\ell^\da\phi_\ell)_{(1')^*}\!+\!h.c.\right]  = \del M_{\phi_e}^2\phi_e^\da\phi_e+\del M_{\phi_\mu}^2 \phi_\mu^\da\phi_\mu+\del M_{\phi_\tau}^2\phi_\tau^\da\phi_\tau
\end{align}
where
\begin{align}\label{eq:flavor dependence of PH mass}
&\del M_{\phi_e}^2 = A\!+\!\re(B\!+\!C)\;,\;\; \del M_{\phi_\mu}^2 = A \!+\! \re(\w B\!+\!\w^*C)\;,\;\; \del M_{\phi_\tau}^2 = A\!+\!\re(\w^*B\!+\!\w C)\;.
\end{align}
The couplings $a_{0,1,2}$, $b_{1,2}$, and $c_{1,2}$ are in general complex and can be adjusted to give different values for each of the $\del M_{\phi_\al}^2$.
\section{Discussion}\label{sec:discussion}
We have presented a group theoretic origin for the leptonic Private Higgs model using the discrete group $\sig = (\zb_3\times\zb_3\times\zb_3)\rx \zb_3$. 
\\\\
The model contains extra Higgs doublets with a common mass at the TeV scale, whose vacuum expectation values determine the charged lepton mass hierarchy $m_e \ll m_\mu \ll m_\tau$. The model also contains three nearly degenerate TeV-scale, gauge-singlet neutrinos and somewhat heavier charged $SU(2)$-singlet bosons. The model relies on the existence of TeV-scale SM-singlet scalars, whose VEVs drive electroweak symmetry breaking for the extra Higgs fields and provide masses for the heavy neutrinos.
\\\\
Given that much of the new physics here is proposed to occur at the TeV scale, the model should be readily falsifiable depending on the results of LHC searches for new physics. In the end of the day, one could always push the masses of the PH fields higher, but one might then have to worry about sizable RG flows in the scalar potential. A full discussion of LHC phenomenology should be done in the context of a $\sig$-based Private Higgs model for quarks, which we postpone to future work.
\\\\
The key assumption is that the scalar potential supports a configuration in which the lepton masses break $\sig$ softly to an abelian discrete subgroup $\zb_3\times\zb_3\times\zb_3$, and we have provided one explicit example of such a configuration. We encourage the reader to find alternative justifications for the soft potential in Eq.~(\ref{eq:softpotential}).
\\\\
Although the $\sig$ symmetry reduces the number of free parameters from the original Private Higgs model, a further reduction of free parameters is desirable. Since $\sig$ is a subgroup of $U(3)$, it is conceivable that this model can be embedded in a higher theory that could provide a dynamical justification for the magnitudes of the various VEVs in the theory.
\\\\
\textit{Acknowledgments:}
\\\\
Part of this work was done while one of us (A. Z.) was visiting the Academia Sinica in Taipei, Republic of China, whose warm hospitality is greatly appreciated. We thank Rafael Porto for early discussions. This research was supported by the NSF under Grant No. PHY07-57035.
\appendix
\section{The group $\Sigma(81)$}\label{sec:appendix}
The group $\sig \equiv (\zb_3\times\zb_3\times\zb_3)\rx\zb_3$ can be defined as the set of diagonal 3-by-3 matrices of the form diag$(\w^{p_1},\w^{p_2},\w^{p_3})$, where $\w \equiv e^{\,i2\pi/3}$ is the cube root of 1 and $p_i$ are integers, supplemented with right-multiplication by the cyclic permutation matrix $c$ and its inverse $a = c^{-1} = c^T$ [see Eq.~(\ref{eq:cyclicgenerators})]. There are $3^3\times3 = 81$ such matrices, so this is a non-abelian discrete group of order 81. In this appendix we detail the construction of this group and derive its equivalence classes and tensor multiplication rules.
\subsection{Preliminaries}
Let $a$ and $b$ be elements of the groups $A$ and $B$, respectively. Let $B$ be a subgroup of $A$, denoted $B<A$. If $aba^{-1}$ is also an element of $B$ (for all $a\;\ep\;A$ and $b\;\ep\;B$) then $B$ is said to be an invariant subgroup\footnote{This is also called a normal subgroup.} of $A$. This is denoted by $B\triangleleft A$.
\\\\
The concept of invariant subgroup can be combined with the concept of product groups to define what is known as the semi-direct product. Let $H$ and $G$ be groups, and define the semi-direct product group $K \equiv H\rx G$. For elements $(h,g)\;\ep\;K$, the semi-direct product is defined by the group multiplication property
\begin{equation}\label{eq:semidirectmult}
(h_1,g_1)\cdot(h_2,g_2) = (h_1 g_1h_2g_1^{-1},g_1 g_2)\;.
\end{equation}
In other words, one uses $G$ to act on elements of $H$ before using those elements of $H$. The operation of Eq.~(\ref{eq:semidirectmult}) makes sense only if $g_1h_2g_1^{-1}\;\ep\;H$, so that elements of the form $(h,I)$ specify an invariant subgroup of $K = H\rx G$.
\\\\
Let us verify some basic group properties of the semi-direct product. First, the group is associative, as can be seen by multiplying three elements in both possible orders:
\begin{align}\label{eq:associativity}
&\left( (h_1,g_1)\cdot(h_2,g_2)\right)\cdot(h_3,g_3) = (h_1g_1h_2g_1^{-1},g_1g_2)\cdot(h_3,g_3) \nonumber\\
&\qquad=(h_1g_1h_2g_1^{-1}(g_1g_2)h_3(g_1g_2)^{-1},g_1g_2g_3) = (h_1g_1h_2g_2h_3g_2^{-1}g_1^{-1},g_1g_2g_3)\nonumber\\
&\nonumber\\
&(h_1,g_1)\cdot\left( (h_2,g_2)\cdot(h_3,g_3)\right) = (h_1,g_1)\cdot(h_2g_2h_3g_2^{-1},g_2g_3) \nonumber\\
&\qquad=(h_1g_1(h_2g_2h_3g_2^{-1})g_1^{-1},g_1g_2g_3) = (h_1g_1h_2g_2h_3g_2^{-1}g_1^{-1},g_1g_2g_3)\;.
\end{align}
Next, let us verify that every element has an inverse by solving the equation $(h_1,g_1)\cdot(h_2,g_2) = (I,I)$ for $(h_2,g_2)$. By Eq.~(\ref{eq:semidirectmult}), we have $g_1g_2 = I \implies g_2 = g_1^{-1}$, and $h_1g_1h_2g_1^{-1} = I \implies h_2 = g_1^{-1}h_1^{-1}g_1$. The latter makes sense because, as stated previously, if $g$ is an element of $G$ and $h$ is an element of $H$, then $ghg^{-1}$ and $g^{-1}h^{-1}g = (ghg^{-1})^{-1}$ are also elements of $H$. Therefore,
\begin{equation}\label{eq:groupinverse}
(h_1,g_1)\cdot(g_1^{-1}h_1^{-1}g_1,g_1^{-1}) = (I,I)\;.
\end{equation}
Furthermore, we can verify that the left inverse equals the right inverse:
\begin{align}
(g_1^{-1}h_1^{-1}&g_1,g_1^{-1})\cdot(h_1,g_1) = (g_1^{-1}h_1^{-1}g_1g_1^{-1}h_1(g_1^{-1})^{-1},g_1^{-1}g_1) =(I,I)\;.
\end{align}
Thus $K = H \rx G$ is a group.
\subsection{$\zb_3\times\zb_3\times\zb_3$}
The cyclic group of order $n$, denoted $\zb_n$, is an abelian group generated by the phase $\w_n \equiv e^{\,i2\pi/n}$. The direct product of $m$ such cyclic groups, $\zb_n^{\,m} \equiv \zb_n\times...\times\zb_n$ ($m$ copies), is an abelian group generated by the elements 
\begin{equation}\label{eq:elements of Z_n^m}
\z_n^{(p_1,p_2,...,p_m)} \equiv (\w_n^{p_1},\w_n^{p_2},...,\w_n^{p_m})
\end{equation}
where the $p_i$ denote the power to which each phase $\w_n$ is raised. The group $\zb_n^{\,m}$ therefore has a total of $n^m$ distinct elements.
\\\\
It is possible to represent the group elements $\z_n^{(p_1,p_2,...,p_m)}$ as diagonal $m$-by-$m$ matrices
\begin{equation}\label{eq:representation of Z_n^m}
R(\z_n^{(p_1,p_2,...,p_m)}) \equiv \text{diag}(\w_n^{p_1},\w_n^{p_2},...,\w_n^{p_m})\;.
\end{equation}
This is just a trivial repackaging of Eq.~(\ref{eq:elements of Z_n^m}), stating that each $\zb_n$ factor lives in its own world. 
\\\\
We now specialize to the case $n = m = 3$. The group $\zb_3$ is generated by the phase $\w \equiv \w_3 \equiv e^{\,i2\pi/3}$, which satisfies $\w^3 = 1$, $\w^2 = \w^*$, and $1+\w+\w^2 = 0$. The 3-by-3 diagonal matrices
\begin{equation}\label{eq:Z3generators}
z_e \equiv R(\z_3^{(1,0,0)}) =\! \ml \w&0&0\\0&1&0\\0&0&1 \mr\!,\; z_\mu \equiv R(\z_3^{(0,1,0)}) = \!\ml 1&0&0\\0&\w&0\\0&0&1 \mr\!,\; z_\tau \equiv R(\z_3^{(0,0,1)}) =\! \ml 1&0&0\\0&1&0\\0&0&\w \mr
\end{equation}
and their inverses generate three copies of $\zb_3$:
\begin{equation}
\zb_3^e \equiv \{I,z_e,z_e^*\}\;,\;\; \zb_3^\mu \equiv \{I,z_\mu,z_\mu^*\}\;,\;\; \zb_3^\tau \equiv \{I,z_\tau,z_\tau^*\}\;.
\end{equation}
The matrix products of $z_e$, $z_\mu$, and $z_\tau$ generate the 3-by-3 matrix representation of the direct product group
\begin{equation}
F \equiv \zb_3^e\times\zb_3^\mu\times\zb_3^\tau
\end{equation}
as described by the general case of Eq.~(\ref{eq:representation of Z_n^m}). The $3^3 = 27$ elements of $F$ are:
\begin{align}\label{eq:Z3cubed}
F &= \{I;z_e,z_\mu,z_\tau;z_e^*,z_\mu^*,z_\tau^*; \nonumber\\
&z_ez_\mu,z_\mu z_\tau,z_\tau z_e; z_e^*z_\mu^*,z_\mu^*z_\tau^*,z_\tau^*z_e^*; \nonumber\\
&z_e^*z_\mu,z_\mu^*z_\tau,z_\tau^*z_e; z_ez_\mu,z_\mu z_\tau^*,z_\tau z_e^*; \nonumber\\
&z_e^*z_\mu z_\tau, z_\mu^*z_\tau z_e, z_\tau^*z_ez_\mu; z_e z_\mu^*z_\tau^*,z_\mu z_\tau^*z_e^*,z_\tau z_e^*z_\mu^*; \nonumber\\
&z_ez_\mu z_\tau = \w I, z_e^*z_\mu^*z_\tau^* = \w^*I \}
\end{align}
We emphasize to the reader that we use the compact notation of Eq.~(\ref{eq:Z3generators}) to denote the 3-by-3 direct sum representation for the elements of the group $F$. We could instead use the group elements given in Eq.~(\ref{eq:elements of Z_n^m}). For example, the equation $z_ez_\mu z_\tau = \w I$ could also be written as $\z_3^{(1,0,0)}\z_3^{(0,1,0)}\z_3^{(0,0,1)}=(\w,\w,\w) = \w (1,1,1) = \w I$. In slight abuse of notation, the symbol $I$ should be understood as the identity element for whichever representation is employed. 
\\\\
At this stage the matrix representation is just a trivial repackaging of the same information, namely that the different $\zb_3^\al$ live in different spaces and therefore do not talk to each other. However, it is possible to extend the construction such that the different spaces get mixed together, and for that purpose the matrix representation will prove convenient.
\subsection{$\Sigma(81) = (\zb_3^e\times\zb_3^\mu\times\zb_3^\tau)\rx\zb_3^C$}
The 3-by-3 representation of $\zb_3\times\zb_3\times\zb_3$ introduced in the previous section suggests an extension that permutes the three $\zb_3$ groups.
\\\\
Consider a fourth $\zb_3$ cyclic group whose elements are written in the (reducible) 3-dimensional representation\footnote{This representation is reducible as follows. Let a vector $(v_1,v_2,v_3)$ transform under the 3-dimensional representation of $\zb_3$. Then the combination $v_1+v_2+v_3$ is invariant under $\zb_3$, the combination $v_1+\w\,v_2+\w^*v_3$ transforms with the phase $\w$ under $c$, and the combination $v_1+\w^*v_2+\w\,v_3$ transforms with the opposite phase $\w^*$ under $c$. Thus the triplet reduces into three singlets: $3 = 1\+1'\+(1')^*$.}:
\begin{equation}\label{eq:cyclicgenerators}
\zb_3^{C} = \left\{I = \ml 1&0&0\\0&1&0\\0&0&1 \mr\;,\;\; c = \ml 0&0&1\\1&0&0\\0&1&0 \mr\;,\;\; a = \ml 0&1&0\\0&0&1\\1&0&0 \mr  \right\}\;.
\end{equation}
Observe that the elements of $\zb_3^{C}$ transform elements of $F$ into other elements of $F$, due to the properties
\begin{equation}\label{eq:cyclegenerators}
c\ml z_e\\ z_\mu\\ z_\tau \mr c^{-1} = \ml z_\mu\\ z_\tau\\ z_e \mr\;,\;\; a\ml z_e\\ z_\mu\\ z_\tau \mr a^{-1} = \ml z_\tau\\ z_e\\ z_\mu \mr\;.
\end{equation}
This allows us to define the semi-direct product group\footnote{This group is of the form $\Sigma(3N^3) \equiv (\zb_N\times\zb_N\times\zb_N)\rx \zb_3$ with $N = 3$, hence the name $\Sigma(81)$.}
\begin{equation}\label{eq:wreathproductgroup}
\Sigma(81) \equiv (\zb_3^e\times\zb_3^\mu\times\zb_3^\tau)\rx \zb_3^{C}\;.
\end{equation}
Groups of the form $H^n\rx \Pi_n$, where $H$ is any finite group and $\Pi_n$ is either $S_n$ or a subgroup of $S_n$, are called wreath products of $H$ with $\Pi_n$ \cite{wreath}. The group of Eq.~({\ref{eq:wreathproductgroup}}) is therefore called the wreath product of $\zb_3$ with $\zb_3$, and it is a non-abelian finite group of order $3^4 = 81$. Due to Eq.~(\ref{eq:cyclegenerators}), we see that elements of the form $(\z,I)$, where $\z\;\ep\;F=\zb_3^e\times\zb_3^\mu\times\zb_3^\tau$, and where $I$ is the identity element of $\zb_3^{C}$, form an invariant subgroup of $\Sigma(81)$. Note that the elements of $\sig$ are unitary but have determinant not necessarily equal to 1, meaning that $\sig$ is a subgroup of $U(3)$ but not of $SU(3)$.
\\\\
For an example of the group multiplication rule, consider $z_e,z_\tau\;\ep\;F$ and $c,a\;\ep\;\zb_3^{C}$. Then the action of $(z_e,c)\;\ep\;\sig$ on $(z_\tau^*,a)\;\ep\;\sig$ is given by
\begin{equation}\label{eq:examplemult}
(z_e,c)\cdot (z_\tau^*,a) = (z_e (cz_\tau^* c^{-1}),c a) = (z_ez_e^*,c\,c^{-1}) = (I,I)\;\ep\;\sig\;.
\end{equation}
This also shows that $(z_\tau^*,a)$ is the inverse of $(z_e,c)$ in $\sig$.
\subsection{Equivalence classes}
It is now possible to separate the elements of $\sig$ into equivalence classes by explicitly performing similarity transformations on all of its 81 elements. 
\\\\
To do this we will employ the matrices of Eqs.~(\ref{eq:Z3generators}) and~(\ref{eq:cyclicgenerators}), which constitute the defining three-dimensional irreducible representation of $\Sigma(81)$, which we denote simply by 3. Let $\z\;\ep\;\zb_3^e\times\zb_3^\mu\times\zb_3^\tau$ and $\s\;\ep\;\zb_3^C$. Then the elements $(\z,\s)\;\ep\;\Sigma(81)$ act on a triplet $\ph \sim 3$ by the ordered matrix multiplication given by
\begin{equation}\label{eq:definingmult}
(\z,\s):\qquad \ph \to \ph\,' = \z\s\,\ph\;.
\end{equation}
As discussed, the matrices $\z$ and $\s$ do not commute, and their order of multiplication is specified by the general multiplication rule of Eq.~(\ref{eq:semidirectmult}) and the example of Eq.~(\ref{eq:examplemult}).\footnote{This is a matter of convention, since $\zeta\s = \s\s^{-1}\z\s = \s\z'$, where $\z' \equiv \s^{-1}\z\s$ is also an element of $F = \zb_3^e\times\zb_3^\mu\times\zb_3^\tau$ since $F$ is an invariant subgroup of $\Sigma(81) = F\rx \zb_3^C$.}
\\\\
To separate the elements of $\Sigma(81)$ into equivalence classes, it will be sufficient to consider similarity transformations of the form $g \to g_\al^c\,g\,(g_\al^c)^{-1}$, where $g_\al^c \equiv z_\al c$ and $g\;\ep\;\Sigma(81)$. Defining $g_\al^a \equiv z_\al a$ and using Eq.~(\ref{eq:cyclegenerators}), the group inverses of $g_\al^c$ and $g_\al^a$ are found to be
\begin{equation}\label{eq:groupinverses}
(g_\al^c)^{-1} = (g_{\al-1}^a)^*\qquad\text{ and }\qquad(g_\al^a)^{-1} = (g_{\al+1}^c)^*\;.
\end{equation}
For the particular elements $(\z,\s) = (z_\beta, I) = z_\beta$ [the last equality denotes the representation specified by Eq.~(\ref{eq:definingmult})], with $\beta = e,\mu,\tau$, we have
\begin{equation}\label{eq:grouparithmetic}
g_\al^cz_\beta (g_\al^c)^{-1} = g_\al^c\,z_\beta(g_{\al-1}^a)^* = z_\al c z_\beta z_{\al-1}^*a = z_\al cz_\beta c^{-1}cz_{\al-1}^*c^{-1} = z_\al z_{\beta+1}z_\al^*= z_{\beta+1}\;.
\end{equation}
The elements $z_e,z_\mu,z_\tau$ cycle into each other and form an equivalence class $C_\al$ with character $\chi(C_\al) = \tr(z_e) = 2+\w$. The conjugate elements $z_e^*,z_\mu^*,z_\tau^*$ form another class $C^*_\al$ with character $\chi(C_\al^*) = 2+\w^*$. 
\\\\
The other classes, to be listed below, are deduced as follows.
\\\\
Applying the group arithmetic of Eq.~(\ref{eq:grouparithmetic}) to elements of the form $z_\al z_\beta$ ($\al\neq\beta$), we find that they form a class $C_{\al\beta}$ with character $\chi(C_{\al\beta}) = \tr(z_ez_\mu) = 1+2\w$. Their conjugates $z_\al^*z_\beta^*$ form a class $C_{\al\beta}^*$ with character $\chi(C_{\al\beta}^*) = 1+2\w^*$. 
\\\\
Furthermore, elements of the form $z_\al z_\beta^*$ [where $(\al,\beta) = (e,\mu),\,(\mu,\tau),\,(\tau,e)$] form a class $D_\al^{\;\;\beta}$ with character $\chi(D_\al^{\;\;\beta}) = \tr(z_ez_\mu^*) = 1+\w+\w^* = 0$. Their conjugates $z_\al^*z_\beta$ form a separate class $(D_\al^{\;\;\beta})^* $ with character also equal to zero.
\\\\
Next we have elements of the form $z_\al z_\beta z_\g^*$ [where $(\al,\beta,\g) = (e,\mu,\tau)$, $(\tau,e,\mu)$, $(\mu,\tau,e)$], which form a class $E_{\al\beta}^\g$ with character $\chi(E_{\al\beta}^\g) = \tr(z_ez_\mu z_\tau^*) = 2\w+\w^*$. The conjugates form a separate class $E^{\al\beta}_\g = (E^\g_{\al\beta})^*$ with character $\chi(E^{\al\beta}_\g) = 2\w^*+\w$.
\\\\
We also have the element $z_ez_\mu z_\tau$, which by itself forms a class $F$ with character $\chi (F) = \tr(z_ez_\mu z_\tau) = 3\w$. The conjugate element $z_e^*z_\mu^*z_\tau^*$ forms a class $F^*$ with character $\chi(F^*) = 3\w^*$.
\\\\
The remaining classes are specified by elements defined with the cyclic permutation matrix $c$ and its inverse $a = c^{-1} = c^2$. These classes will always have character equal to zero, since the $\z\;\ep\;\zb_3^e\times\zb_3^\mu\times\zb_3^\tau$ are diagonal, while $c$ and $a$ are strictly off-diagonal.
\\\\
Consider using the element $g_\al^c$ to perform a similarity transformation on another element $g_\beta^c$. Using Eqs.~(\ref{eq:cyclegenerators}) and~(\ref{eq:groupinverses}), we have:
\begin{equation}
g_\al^c\,g_\beta^c\,(g_\al^c)^{-1} = g_\al^cg_\beta^c(g_{\al-1}^a)^* = z_\al cz_\beta c z_{\al-1}^* a = z_\al cz_\beta c^{-1}a z_{\al-1}^*a^{-1}c = z_\al z_{\beta+1}z_{\al-2}^*c\;.
\end{equation}
For example, $g_e^c g_\mu^c(g_e^c)^{-1} = z_ez_\tau z_\mu^*c$. So elements of the form $z_\al c$ and $z_\al z_\beta z_\g^*c$ belong to the same class, which we call $C_\al^c$. Their conjugates form a separate class, $(C_\al^{c})^*$. 
\\\\
Observe that elements of these classes do not cube to the identity. For concreteness, consider the element $z_ec$:
\begin{equation}
(z_ec)^3 \!=\! z_ec\,z_ec\,z_ec = z_ea^{-1}z_e c\, z_e c = a^{-1}(a z_e a^{-1})z_e(c z_e c^{-1})c^2= a^{-1}z_\tau z_e z_\mu c^{-1}= c\, \w I c^{-1} = \w I\;.
\end{equation}
Therefore, the minimum power to which elements in the classes $C_\al^c$ and $(C_\al^c)^*$ must be raised to obtain the identity is six.
\\\\
Next, perform the similarity transformation by $g_e^c$ on $z_\mu z_\tau c$:
\begin{equation}
g_e^c(z_\mu z_\tau c)(g_e^c)^{-1} = z_ecz_\mu z_\tau c z_\tau^*a = z_e(cz_\mu c^{-1})(cz_\tau c^{-1})(az_\tau^*a^{-1})c = z_e^*z_\mu^*z_\tau c\;.
\end{equation}
Therefore the elements $z_ez_\mu c, z_\mu z_\tau c, z_\tau z_e c$ and $z_e^*z_\mu^*z_\tau c,z_\mu^*z_\tau^*z_e c,z_\tau^*z_e^*z_\mu c$ belong to the same class. But since $z_ez_\tau z_\mu^* = (z_e^*z_\tau^*z_\mu)^*$ belongs to the same class as $z_\mu c = (z_\mu^*c)^*$, these elements belong to the classes $C_\al^c$ and $(C_\al^{c})^{*}$ defined previously. 
\\\\
Next, consider the similarity transformation by $g_e^c$ of the element $z_\mu z_\tau^*c$:
\begin{equation}
g_e^c(z_\mu z_\tau^*c)(g_e^c)^{-1} = z_ecz_\mu z_\tau^*cz_\tau^*a = z_e(cz_\mu c^{-1})(cz_\tau^*c^{-1})(a z_\tau^*a^{-1})c = z_ez_\tau z_e^*z_\mu^*c = z_\tau z_\mu^*c\;.
\end{equation}
We see that $z_\mu z_\tau^*c$ and $z_\mu^*z_\tau c = (z_\mu z_\tau^*c)^*$ are related by a similarity transformation and therefore belong to the same class. The elements $z_ez_\mu^* c,z_\mu z_\tau^* c,z_\tau z_e^* c$ and their conjugates form a self-conjugate class $D_\al^{c\;\;\beta} = (D_\al^{c\;\;\beta})^*$. These elements cube to the identity:
\begin{align}
(z_\mu z_\tau^* c)^3 &= z_\mu z_\tau^* c\,z_\mu z_\tau^* c\,z_\mu z_\tau^* c= c(az_\mu a^{-1})(az_\tau^*a^{-1})z_\mu z_\tau^*(cz_\mu c^{-1})(cz_\tau^*c^{-1})c^2 \nonumber\\
& = c(z_e z_\mu^*)z_\mu z_\tau^*(z_\tau z_e^*) c^{-1} = I\;.
\end{align}
Observe that
\begin{equation}
g_e^c(z_e^*z_\mu^*z_\tau^*c)(g_e^c)^{-1} = z_ecz_e^*z_\mu^*z_\tau^*cz_\tau^*a = z_e(cz_e^*c^{-1})(cz_\mu^*c^{-1})(cz_\tau^*c^{-1})(az_\tau^*a^{-1})c = z_\mu z_\tau^*c\;.
\end{equation}
Therefore $z_e^*z_\mu^*z_\tau^*c$ belongs to the same class as $z_\mu z_\tau^*c$. Additionally, the element $c$ by itself belongs to this class, since $z_ecz_e^{-1} = z_ecz_e^{-1}c^{-1}c = z_ez_\mu^{-1}c = z_ez_\mu^*c$.
\\\\
The above elements but with $c$ replaced by $a$ form their own classes, following the same procedure as above. Thus we also have the classes $C_{\al}^a$, $(C_{\al}^{a})^{*}$, and $D_\al^{a\;\;\beta} = (D_\al^{a\;\;\beta})^*$. 
\\\\
In total, including the identity element as its own class $C_I$, we have:
\begin{equation}
C_I,C_\al,C_\al^*,C_{\al\beta},C_{\al\beta}^*,D_\al^{\;\;\beta},(D_\al^{\;\;\beta})^*,E^\g_{\al\beta}, (E^\g_{\al\beta})^*,F,F^*,C_\al^c,(C_\al^{c})^{*}, D_\al^{c\;\;\beta},C_\al^a,(C_\al^{a})^{*}, D_\al^{a\;\;\beta}\;.
\end{equation}
This makes a total of $n_C = 17$ equivalence classes.
\subsection{Irreducible representations}
We now use the defining triplet representation to find other irreducible representations.
\\\\
Let $\ph = (\ph_e,\ph_\mu,\ph_\tau)$ be a field that transforms under the defining triplet representation, and consider the group element $g = (z_e,I)$. According to Eq.~(\ref{eq:definingmult}), the action of $g$ on $\ph$ is given by $(z_e,I):\ph \to z_e\ph$, or in components
\begin{equation}\label{eq:(z_e,I)on3}
(z_e,I): \ml \ph_e\\ \ph_\mu\\ \ph_\tau \mr \to \ml \w\,\ph_e\\ \ph_\mu\\ \ph_\tau \mr\;.
\end{equation}
Letting $\chi = (\chi_e,\chi_\mu,\chi_\tau) \sim 3$ be another triplet field, we can use Eq.~(\ref{eq:(z_e,I)on3}) to decompose the product $\ph\chi \sim 3\x3$ into irreducible representations. We have:
\begin{equation}\label{eq:(z_e,I)on3'}
(z_e,I): \ml \ph_e\chi_e\\ \ph_\mu\chi_\mu\\ \ph_\tau\chi_\tau \mr \to \ml \w^2\ph_e\chi_e\\ \ph_\mu\chi_\mu\\ \ph_\tau\chi_\tau \mr\;,\;\; \ml \ph_\mu\chi_\tau\\ \ph_\tau \chi_e\\ \ph_e\chi_\mu \mr \to \ml \ph_\mu\chi_\tau\\ \w\ph_\tau\chi_e \\ \w\ph_e\chi_\mu\mr\;,\;\; \ml \ph_\tau\chi_\mu\\ \ph_e\chi_\tau\\ \ph_\mu\chi_e \mr \to \ml \ph_\tau\chi_\mu\\ \w\ph_e\chi_\tau\\ \w\ph_\mu\chi_e \mr\;.
\end{equation}
Since $\w^2 = \w^*$, the first of these transforms as the conjugate $3^*$ of the defining representation. The second transforms differently from $3$ and $3^*$, and therefore defines a new triplet representation, $3'$. The third transforms in the same way as the second, and so also transforms under $3'$. 
\\\\
Thus we arrive at the multiplication rule
\begin{equation}\label{eq:3x3}
3\x3 = 3^*\+3'_1\+3'_2
\end{equation}
where, in terms of the components of $\ph \sim 3$ and $\chi \sim 3$, this reads explicitly
\begin{equation}\label{eq:3x3components}
\ml \ph_e\\ \ph_\mu\\ \ph_\tau \mr \x \ml \chi_e\\ \chi_\mu\\ \chi_\tau \mr = \ml \ph_e\chi_e\\ \ph_\mu\chi_\mu\\ \ph_\tau\chi_\tau \mr \+ \ml \ph_\mu\chi_\tau\\ \ph_\tau\chi_e\\ \ph_e\chi_\mu \mr \+ \ml \ph_\tau\chi_\mu\\ \ph_e\chi_\tau\\ \ph_\mu\chi_e\\ \mr\;.
\end{equation}
If desired, it is possible to change basis by symmetrizing and antisymmetrizing the new triplets $3'_i$, in which case we write $3'_1\+3'_2 = 3'_A\+3'_S$.
\\\\
Next consider the product $\ph\chi^* \sim 3\x3^*$. Using Eq.~(\ref{eq:(z_e,I)on3}) and its conjugate, we find
\begin{equation}\label{eq:3x3*under(z_e,I)}
(z_e,I):\ml \ph_e\chi_e^*\\ \ph_\mu\chi_\mu^*\\ \ph_\tau\chi_\tau^* \mr \to \ml \ph_e\chi_e^*\\ \ph_\mu\chi_\mu^*\\ \ph_\tau\chi_\tau^* \mr\;,\;\;\ml \ph_\tau\chi_\mu^*\\ \ph_e\chi_\tau^* \\ \ph_\mu\chi_e^*\mr \to \ml \ph_\tau\chi_\mu^*\\ \w\,\ph_e\chi_\tau^*\\ \w^*\ph_\mu\chi_e^* \mr\;,\;\; \ml  \ph_\mu\chi_\tau^*\\ \ph_\tau\chi_e^*\\ \ph_e\chi_\mu^* \mr \to \ml\ph_\mu\chi_\tau^*\\ \w^*\ph_\tau\chi_e^*\\  \w\,\ph_e\chi_\mu^* \mr\;.
\end{equation}
The first triplet is invariant under the $z_\al$ and is therefore reducible under the generators of $\zb_3^C = \{I,c,a\}$ using the usual decomposition $3 \to 1\+1'\+(1')^*$. The second triplet transforms differently from $3,3^*,3'$, and $(3')^*$, and therefore defines another irreducible triplet, $\hat 3$. The third triplet transforms under the conjugate representation, $\hat3^*$. 
\\\\
Thus we arrive at the multiplication rule
\begin{equation}\label{eq:3x3*}
3\x3^* = 1\+1'\+(1')^*\+\hat 3\+\hat3^*
\end{equation}
given in components by
\begin{equation}\label{eq:3x3*components}
\ml \ph_e\\ \ph_\mu\\ \ph_\tau \mr\x\ml \chi_e^*\\ \chi_\mu^*\\ \chi_\tau^* \mr = \left[\begin{matrix} \ph_e\chi_e^*+\ph_\mu\chi_\mu^*+\ph_\tau\chi_\tau^* \\ \ph_e\chi_e^*+\w\,\ph_\mu\chi_\mu^*+\w^*\ph_\tau\chi_\tau^* \\ \ph_e\chi_e^*+\w^*\ph_\mu\chi_\mu^*+\w\,\ph_\tau\chi_\tau^*\end{matrix}\right] \+\ml \ph_\tau\chi_\mu^*\\ \ph_e\chi_\tau^*\\ \ph_\mu\chi_e^* \mr\+ \ml \ph_\mu\chi_\tau^*\\ \ph_\tau\chi_e^*\\  \ph_e\chi_\mu^* \mr\;.
\end{equation}
The square brackets denote that the three components transform reducibly as distinct singlet representations, as discussed. 
\\\\
Next let $\ph \sim 3$ as before, but introduce a triplet $\psi \sim 3'$ that transforms under $(z_e,I)$ as specified by Eq.~(\ref{eq:(z_e,I)on3'}): $\psi \to \text{diag}(1,\w,\w)\psi = z_\mu z_\tau\psi$. Forming the product $\ph\,\psi \sim 3\x3'$, we find:
\begin{equation}\label{eq:3times3'under(z_e,I)}
(z_e,I):\ml \ph_e\psi_e\\ \ph_\mu\psi_\mu\\ \ph_\tau\psi_\tau \mr \to \ml \w\,\ph_e\psi_e\\ \w\,\ph_\mu\psi_\mu\\ \w\,\ph_\tau\psi_\tau \mr\;,\;\;\ml \ph_\tau\psi_e\\ \ph_e\psi_\mu\\ \ph_\mu\psi_\tau \mr \to \ml \ph_\tau\psi_e\\ \w^*\ph_e\psi_\mu\\ \w\,\ph_\mu\psi_\tau \mr\;,\;\;\ml \ph_\mu\psi_e\\ \ph_\tau\psi_\mu\\ \ph_e\psi_\tau \mr \to \ml \ph_\mu\psi_e\\ \w\,\ph_\tau\psi_\mu\\ \w^*\ph_e\psi_\tau \mr\;.
\end{equation}
Comparing to Eq.~(\ref{eq:3x3*under(z_e,I)}), we see that the second and third triplets of Eq.~(\ref{eq:3times3'under(z_e,I)}) transform as $\hat3^*$ and $\hat3$, respectively. 
\\\\
The first triplet of Eq.~(\ref{eq:3times3'under(z_e,I)}) transforms differently from all of the triplets considered previously. Since all components transform with the same phase, this triplet is reducible into singlets. Moreover, since these singlets transform nontrivially under the action of the $(z_\al, I)$, indicating that they are distinct from the singlets $1'$ and $(1')^*$ of Eq.~(\ref{eq:3x3*}).
\\\\
Thus we have the multiplication rule
\begin{equation}\label{eq:3x3'}
3\x3' = \widetilde 1\+\widetilde 1'\+\widetilde 1''\+\hat3\+\hat3^*
\end{equation}
given in components by
\begin{equation}\label{eq:3x3'components}
\ml \ph_e\\ \ph_\mu\\ \ph_\tau \mr\x\ml \psi_e\\ \psi_\mu\\ \psi_\tau \mr = \left[ \begin{matrix}   \ph_e\psi_e+\ph_\mu\psi_\mu+\ph_\tau\psi_\tau\\ \ph_e\psi_e+\w\,\ph_\mu\psi_\mu+\w^*\ph_\tau\psi_\tau\\  \ph_e\psi_e+\w^*\ph_\mu\psi_\mu+\w\,\ph_\tau\psi_\tau\end{matrix}\right]\+\ml \ph_\mu\psi_e\\ \ph_\tau\psi_\mu\\ \ph_e\psi_\tau \mr\+\ml \ph_e\psi_e\\ \ph_\mu\psi_\mu\\ \ph_\tau\psi_\tau \mr\;.
\end{equation}
As for Eq.~(\ref{eq:3x3*components}), the square brackets in Eq.~(\ref{eq:3x3'components}) indicate that the first triplet is reducible as follows. The singlet $\ph_e\psi_e+\ph_\mu\psi_\mu+\ph_\tau\psi_\tau \sim \widetilde 1$ transforms with $\w$ under $(z_\al,I)$ but is invariant under $(I,c)$. The singlet $\ph_e\psi_e+\w\,\ph_\mu\psi_\mu+\w^*\ph_\tau\psi_\tau \sim \widetilde 1'$ transforms with the phase $\w$ under both $(z_\al,I)$ and $(I,c)$. The singlet $\ph_e\psi_e+\w^*\ph_\mu\psi_\mu+\w\,\ph_\tau\psi_\tau\sim \widetilde1''$ transforms with the phase $\w$ under $(z_\al,I)$ but with the opposite phase $\w^*$ under $(I,c)$.
\\\\
Each of these non-invariant singlets is complex, so overall we have $1',\widetilde 1,\widetilde 1',\widetilde 1''$, and their conjugates, which along with the invariant $1$ makes a total of 9 distinct singlets. The group elements acting on the these representations are given in Table~\ref{table:generators on 1}.
\begin{table}[h]
\centering
\fbox{\begin{minipage}{14cm}
	\centering
	\begin{tabular}{l ||*{2}{c|}}
          $\Sigma(81)$   & $(z_\al,I)$ & $(I,c)$ \\
\hline\hline
$1$ & $1$&$1$\\
$1'$ & $1$ & $\w$  \\
$\widetilde 1$ & $\w$ & $1$\\
$\widetilde 1'$ & $\w$ & $\w$ \\
$\widetilde 1''$ & $\w$& $\w^*$
	\end{tabular}
	\caption{Elements of $\Sigma(81)$ acting on the distinct singlet representations. The 1 is invariant under all elements of the group. The other singlets are not invariant under group transformations, and they are complex.}
	\label{table:generators on 1}
\end{minipage}}
\end{table}
\\\\
Let $\psi \sim 3'$ and $\xi \sim \hat3$. The transformation properties under the action of $(z_e,I)$ are given by Eqs.~(\ref{eq:(z_e,I)on3'}) and~(\ref{eq:3x3*under(z_e,I)}) as $\psi \to z_\mu z_\tau\psi$ and $\xi \to z_\mu z_\tau^*\xi$. The components of the product $\psi\xi \sim 3'\x\hat3$ transform as:
\begin{equation}\label{eq:(z_e,I)on3'x3''}
(z_e,I): \ml \psi_\mu\xi_\mu\\ \psi_\tau\xi_\tau\\ \psi_e\xi_e \mr \to \ml \w^*\psi_\mu\xi_\mu\\ \psi_\tau\xi_\tau\\ \psi_e\xi_e \mr\;,\;\; \ml \psi_\mu\xi_\tau\\ \psi_\tau \xi_e\\ \psi_e\xi_\mu \mr \to \ml \psi_\mu\xi_\tau\\ \w\,\psi_\tau\xi_e\\ \w\,\psi_e\xi_\mu \mr\;,\;\; \ml \psi_\mu\xi_e\\ \psi_\tau \xi_\mu\\ \psi_e\xi_\tau \mr \to \ml \w\,\psi_\mu\xi_e\\ \w^*\psi_\tau\xi_\mu\\ \w^*\psi_e\xi_\tau \mr\;.
\end{equation}
The first triplet of Eq.~(\ref{eq:(z_e,I)on3'x3''}) transforms as $3^*$, and the second as $3'$. The third transforms differently from $3,3',\hat3$ or their conjugates, and therefore specifies a new complex irreducible triplet, $\tilde 3$.
\\\\
Therefore we find the multiplication rule
\begin{equation}\label{eq:3'x3''}
3'\x \hat3 = 3^*\+3'\+\tilde 3
\end{equation}
which in components reads
\begin{equation}\label{eq:3'x3''components}
\ml \psi_e\\ \psi_\mu\\ \psi_\tau \mr\x\ml \xi_e\\ \xi_\mu\\ \xi_\tau \mr = \ml \psi_\mu\xi_\mu\\ \psi_\tau\xi_\tau\\ \psi_e\xi_e \mr\+\ml \psi_\mu\xi_\tau\\ \psi_\tau\xi_e\\ \psi_e\xi_\mu \mr\+\ml \psi_\mu\xi_e\\ \psi_\tau\xi_\mu\\ \psi_e\xi_\tau \mr\;.
\end{equation}
Thus far we have found the irreducible triplets $3,3',\hat3,\tilde3$, and their conjugates, making a total of 8 distinct triplets. The group elements acting on these triplets is given in Table~\ref{table:generators on 3}.
\\
\begin{table}[h]
\centering
\fbox{\begin{minipage}{14cm}
	\centering
	\begin{tabular}{l ||*{3}{c|}}
          $\Sigma(81)$   & $(z_e,I)$ & $(z_\mu,I)$ & $(z_\tau,I)$ \\
\hline\hline
$3$ & $z_e$&$z_\mu$&$z_\tau$\\
$3'$ & $z_\mu z_\tau$ & $z_\tau z_e$ & $z_e z_\mu$  \\
$\hat 3$ & $z_\mu z_\tau^*$ & $z_\tau z_e^*$ & $z_e z_\mu^*$\\
$\tilde 3$ & $z_ez_\mu^* z_\tau^*$ & $z_\mu z_\tau^*z_e^*$ & $z_\tau z_e^*z_\mu^*$ 
	\end{tabular}
	\caption{Elements of $\Sigma(81)$ acting on the irreducible complex triplet representations. The representation denoted by 3 is the defining representation, whose 3-by-3 matrices were used to derive the properties of the group.}
	\label{table:generators on 3}
\end{minipage}}
\end{table}
\\\\
A convenient way to denote the content of Table 2 is to label the components of the four triplets as follows:
\begin{align}\label{eq:triplet notation}
&\phi \sim3 \implies \phi = (\phi_e,\phi_\mu,\phi_\tau) \nonumber\\
&h \sim 3' \implies h = (h_{\mu\tau},h_{\tau e},h_{e\mu}) \sim (\phi_\mu\phi_\tau,\phi_\tau\phi_e,\phi_e\phi_\mu) \nonumber\\
&\chi \sim \hat3 \implies \chi = (\chi_\tau^{\;\;\mu},\chi_e^{\;\;\tau},\chi_\mu^{\;\;e}) \sim (\phi_\tau\phi_\mu^*,\phi_e\phi_\tau^*,\phi_\mu\phi_e^*)\nonumber\\
&\z \sim \tilde 3 \implies \z = (\z_e^{\;\;\mu\tau},\z_\mu^{\;\;\tau e},\z_\tau^{\;\;e\mu}) \sim (\phi_e\phi_\mu^*\phi_\tau^*,\phi_\mu\phi_\tau^*\phi_e^*,\phi_\tau\phi_e^*\phi_\mu^*)\;.
\end{align}
Fields with one lower index $\al = e,\mu,\tau$ transform under the defining 3-dimensional representation, and fields with one upper index transform under the conjugate $3^*$. Fields with two indices (either two lower, or one lower and one upper) are organized cyclically: $``h_e" \equiv h_{\mu\tau}$, $``h_\mu" \equiv h_{\tau e}$, and $``h_\tau" \equiv h_{e\mu}$. Fields with one lower and two upper (and the conjugate case of one upper and two lower) are organized according to the ``odd index out": $``\z_e" \equiv \z_e^{\;\;\mu\tau}$, $``\z_\mu" \equiv \z_\mu^{\;\;\tau e}$, and $``\z_\tau" \equiv \z_\tau^{\;\;e\mu}$.
\\\\
This notation makes it easy to carry out group multiplication and decompose products into irreducible representations. For example, using the notation of Eq.~(\ref{eq:triplet notation}) and the transformation rules of Tables~\ref{table:generators on 1} and~\ref{table:generators on 3}, we can easily decompose the product of the defining triplet $3$ with itself \footnote{In Eq.~(\ref{eq:3 x 3 = 3*+3'+3'}) the prime on $\phi'$ denotes that it is a different field from $\phi$, but both fields still transform as the defining $3$ representation.} or with $3',\hat3$ and $\tilde 3$:
\begin{align}
&3\x3 \sim \ml \phi_e\\ \phi_\mu\\ \phi_\tau \mr\!\x\!\ml \phi'_e\\ \phi'_\mu\\ \phi'_\tau \mr = \ml \phi_e\phi'_e\\ \phi_\mu\phi'_\mu\\ \phi_\tau\phi'_\tau \mr\+\ml \phi_\mu\phi'_\tau\\ \phi_\tau\phi'_e\\ \phi_\tau\phi'_\mu \mr\+\ml \phi_\tau\phi'_\mu\\ \phi_e\phi'_\tau\\ \phi_\mu\phi'_e \mr \sim 3^*\+3'_1\+3'_2 \label{eq:3 x 3 = 3*+3'+3'}
\end{align}
\begin{align}
&3\x3'\! \sim\! \ml \phi_e\\ \phi_\mu\\ \phi_\tau \mr\!\x\!\ml h_{\mu\tau}\\ h_{\tau e}\\ h_{e\mu} \mr\! =\! \left[ \begin{matrix} \phi_e h_{\mu\tau}+\phi_\mu h_{\tau e}+\phi_\tau h_{e\mu}\\ \phi_e h_{\mu\tau}+\w\,\phi_\mu h_{\tau e}+\w^*\phi_\tau h_{e\mu} \\ \phi_e h_{\mu\tau}+\w^*\phi_\mu h_{\tau e}+\w\,\phi_\tau h_{e\mu} \end{matrix}\right]\+\ml \phi_\mu h_{\mu\tau}\\ \phi_\tau h_{\tau e}\\ \phi_e h_{e\mu} \mr\+\ml \phi_\tau h_{\mu\tau}\\ \phi_e h_{\tau e}\\ \phi_\mu h_{e\mu} \mr \nonumber\\
& \qquad\sim \widetilde 1\+\widetilde 1'\+\widetilde 1''\+\hat3\+\hat3^*
\end{align}
\begin{align}
&3\x \hat3 \sim \ml \phi_e\\ \phi_\mu\\ \phi_\tau \mr\!\x\!\ml \chi_\tau^{\;\;\mu}\\ \chi_e^{\;\;\tau}\\ \chi_\mu^{\;\;e} \mr = \ml \phi_\tau\chi_e^{\;\;\tau}\\ \phi_e\chi_\mu^{\;\;e}\\ \phi_\mu\chi_\tau^{\;\;\mu} \mr\+\ml \phi_\tau\chi_\tau^{\;\;\mu}\\ \phi_e\chi_e^{\;\;\tau}\\ \phi_\mu\chi_\mu^{\;\;e} \mr \+\ml \phi_\tau\chi_\mu^{\;\;e}\\ \phi_e\chi_\tau^{\;\;\mu}\\ \phi_\mu\chi_e^{\;\;\tau} \mr \sim 3\+(3')^*\+\tilde 3^*\label{eq:3x3''=...}
\end{align}
\begin{align}
&3\x \tilde3 \sim \ml \phi_e\\ \phi_\mu\\ \phi_\tau \mr\!\x\!\ml \z_e^{\;\;\mu\tau}\\ \z_\mu^{\;\;\tau e}\\ \z_\tau^{\;\;e\mu} \mr = \left[ \begin{matrix}\phi_e\z_e^{\;\;\mu\tau}+\phi_\mu\z_\mu^{\;\;\tau e}+\phi_\tau\z_\tau^{\;\;e\mu}\\ \phi_e\z_e^{\;\;\mu\tau}+\w\,\phi_\mu\z_\mu^{\;\;\tau e}+\w^*\phi_\tau\z_\tau^{\;\;e\mu}\\ \phi_e\z_e^{\;\;\mu\tau}+\w^*\phi_\mu\z_\mu^{\;\;\tau e}+\w\,\phi_\tau\z_\tau^{\;\;e\mu}\end{matrix}\right]\+\ml \phi_e\z_\tau^{\;\;e\mu}\\ \phi_\mu\z_e^{\;\;\mu\tau}\\ \phi_\tau\z_\mu^{\;\;\tau e} \mr\+\ml \phi_e\z_\mu^{\;\;\tau e}\\ \phi_\mu\z_\tau^{\;\;e\mu}\\ \phi_\tau\z_e^{\;\;\mu\tau} \mr \nonumber\\
&\qquad \sim \widetilde 1^*\+(\widetilde 1')^*\+(\widetilde 1'')^*\+\hat3\+\hat3^*
\end{align}
Similarly, we can decompose the product of $3$ with any of the conjugate triplets:
\begin{align}
&3\x3^* \sim \ml \phi_e\\ \phi_\mu\\ \phi_\tau \mr\!\x\!\ml \phi'^e\\ \phi'^\mu\\ \phi'^\tau \mr = \left[ \begin{matrix} \phi_e\phi'^e+\phi_\mu\phi'^\mu+\phi_\tau\phi'^\tau\\ \phi_e\phi'^e+\w\,\phi_\mu\phi'^\mu+\w^*\phi_\tau\phi'^\tau\\ \phi_e\phi'^e+\w^*\phi_\mu\phi'^\mu+\w\,\phi_\tau\phi'^\tau\end{matrix}\right]\+\ml \phi_\tau\phi'^\mu\\ \phi_e\phi'^\tau\\ \phi_\mu\phi'^e \mr\+\ml \phi_\mu\phi'^\tau\\ \phi_\tau\phi'^e\\ \phi_e\phi'^\mu \mr\nonumber\\ 
&\qquad\sim 1\+1'\+(1')^*\+\hat3\+\hat3^* \label{eq:3 x 3* decomposition}
\end{align}
\begin{align}
&3\x(3')^* \sim \ml \phi_e\\ \phi_\mu\\ \phi_\tau \mr\!\x\!\ml h^{\mu\tau}\\ h^{\tau e}\\ h^{e\mu} \mr = \ml \phi_\mu h^{e\mu}\\ \phi_\tau h^{\mu\tau}\\ \phi_e h^{\tau e} \mr\+\ml \phi_\tau h^{\tau e}\\ \phi_e h^{e\mu}\\ \phi_\mu h^{\mu\tau} \mr\+\ml \phi_e h^{\mu\tau}\\ \phi_\mu h^{\tau e}\\ \phi_\tau h^{e\mu} \mr \sim 3_1\+3_2\+\tilde 3\label{eq:3x(3')*}\\
&\nonumber\\
&3\x \hat3^* \sim \ml \phi_e\\ \phi_\mu\\ \phi_\tau \mr\!\x\!\ml \chi_\mu^{\;\;\tau}\\ \chi_\tau^{\;\;e}\\ \chi_e^{\;\;\mu} \mr = \ml \phi_\mu \chi_e^{\;\;\mu}\\ \phi_\tau\chi_\mu^{\;\;\tau}\\ \phi_e\chi_\tau^{\;\;e} \mr\+\ml \phi_\mu\chi_\mu^{\;\;\tau}\\ \phi_\tau\chi_\tau^{\;\;e}\\ \phi_e\chi_e^{\;\;\mu} \mr\+\ml \phi_\mu\chi_\tau^{\;\;e}\\ \phi_\tau\chi_e^{\;\;\mu}\\ \phi_e\chi_\mu^{\;\;\tau} \mr \sim 3\+(3')^*\+\tilde 3^*\\
&3\x \tilde 3^* \sim \ml \phi_e\\ \phi_\mu\\ \phi_\tau \mr\!\x\!\ml \z^e_{\;\;\mu\tau}\\ \z^\mu_{\;\;\tau e}\\ \z^\tau_{\;\;e\mu} \mr = \ml \phi_e\z^e_{\;\;\mu\tau}\\ \phi_\mu\z^\mu_{\;\;\tau e}\\ \phi_\tau\z^\tau_{\;\;e\mu} \mr\+\ml \phi_\mu\z^\tau_{\;\;e\mu}\\ \phi_\tau\z^e_{\;\;\mu\tau}\\ \phi_e\z^\mu_{\;\;\tau e} \mr\+\ml \phi_\tau\z^\mu_{\;\;\tau e}\\ \phi_e\z^\tau_{\;\;e\mu}\\ \phi_\mu\z^e_{\;\;\mu\tau} \mr \sim 3'\+\tilde3_1\+\tilde3_2 \label{eq:3x(3'')*=...}
\end{align}
Next, we have $3'$ times itself, $\hat 3$, and $3'''$:
\begin{align}
&3'\x3' \sim \ml h_{\mu\tau}\\ h_{\tau e}\\ h_{e\mu} \mr\!\x\!\ml h'_{\mu\tau}\\ h'_{\tau e}\\ h'_{e\mu} \mr = \ml h_{\mu\tau}h'_{\mu\tau}\\ h_{\tau e}h'_{\tau e}\\ h_{e\mu}h'_{e\mu} \mr\+\ml h_{e\mu}h'_{\tau e}\\ h_{\mu\tau}h'_{e\mu}\\ h_{\tau e}h'_{\mu\tau} \mr\+\ml h_{\tau e}h'_{e\mu}\\ h_{e\mu}h'_{\mu\tau}\\ h_{\mu\tau}h'_{\tau e} \mr \sim (3')^*\+\tilde3_1^*\+ \tilde3_2^*\\
&3'\x \hat3 \sim \ml h_{\mu\tau}\\ h_{\tau e}\\ h_{e\mu} \mr\!\x\!\ml \chi_\tau^{\;\;\mu}\\ \chi_e^{\;\;\tau}\\ \chi_\mu^{\;\;e} \mr = \ml h_{\tau e}\chi_e^{\;\;\tau}\\ h_{e\mu}\chi_\mu^{\;\;e}\\ h_{\mu\tau}\chi_\tau^{\;\;\mu} \mr\+\ml h_{\tau e}\chi_\mu^{\;\;e}\\ h_{e\mu}\chi_\tau^{\;\;\mu}\\ h_{\mu\tau}\chi_e^{\;\;\tau} \mr\+\ml h_{\tau e}\chi_\tau^{\;\;\mu}\\ h_{e\mu}\chi_e^{\;\;\tau}\\ h_{\mu\tau}\chi_\mu^{\;\;e} \mr\sim 3^*\+3'\+\tilde3\\
&3'\x \tilde3\sim\ml h_{\mu\tau}\\ h_{\tau e}\\ h_{e\mu} \mr\!\x\!\ml \z_e^{\;\;\mu\tau}\\ \z_\mu^{\;\;\tau e}\\ \z_\tau^{\;\;e\mu} \mr = \ml h_{\mu\tau}\z_e^{\;\;\mu\tau}\\ h_{\tau e}\z_\mu^{\;\;\tau e}\\ h_{e\mu}\z_\tau^{\;\;e\mu} \mr\+\ml h_{\tau e}\z_\tau^{\;\;e\mu}\\ h_{e\mu}\z_e^{\;\;\mu\tau}\\ h_{\mu\tau}\z_\mu^{\;\;\tau e} \mr\+\ml h_{e\mu}\z_\mu^{\;\;\tau e}\\ h_{\mu\tau}\z_\tau^{\;\;e\mu}\\ h_{\tau e}\z_e^{\;\;\mu\tau} \mr \sim 3\+(3'_1)^*\+(3'_2)^*
\end{align}
We also have $3'$ times its conjugate, $\hat3^*$, and $ \tilde 3^*$:
\begin{align}
&3'\x(3')^* \sim\ml h_{\mu\tau}\\ h_{\tau e}\\ h_{e\mu} \mr\!\x\!\ml h'^{\mu\tau}\\ h'^{\tau e}\\ h'^{e\mu} \mr = \left[ \begin{matrix} h_{\mu\tau}h'^{\mu\tau}\!+\!h_{\tau e}h'^{\tau e}\!+\!h_{e\mu}h'^{e\mu}\\ h_{\mu\tau}h'^{\mu\tau}\!+\!\w\,h_{\tau e}h'^{\tau e}\!+\!\w^*h_{e\mu}h'^{e\mu}\\ h_{\mu\tau}h'^{\mu\tau}\!+\!\w^*h_{\tau e}h'^{\tau e}\!+\!\w\,h_{e\mu}h'^{e\mu} \end{matrix}\right]\+\ml h_{\tau e}h'^{e\mu}\\ h_{e\mu}h'^{\mu\tau}\\ h_{\mu\tau}h'^{\tau e} \mr\+\ml h_{e\mu}h'^{\tau e}\\ h_{\mu\tau}h'^{e\mu}\\ h_{\tau e}h'^{\mu\tau} \mr \nonumber\\
&\qquad\sim 1\+1'\+(1')^*\+\hat3\+\hat3^*\\
&\nonumber\\
&3'\x \hat3^* \sim \ml h_{\mu\tau}\\ h_{\tau e}\\ h_{e\mu} \mr\!\x\!\ml \chi_\mu^{\;\;\tau}\\ \chi_\tau^{\;\;e}\\ \chi_e^{\;\;\mu} \mr = \ml h_{e\mu}\chi_e^{\;\;\mu}\\ h_{\mu\tau}\chi_\mu^{\;\;\tau}\\ h_{\tau e}\chi_\tau^{\;\;e} \mr\+\ml h_{e\mu}\chi_\tau^{\;\;e}\\ h_{\mu\tau}\chi_e^{\;\;\mu}\\ h_{\tau e}\chi_\mu^{\;\;\tau} \mr\+\ml h_{e\mu}\chi_\mu^{\;\;\tau}\\ h_{\mu\tau}\chi_\tau^{\;\;e}\\ h_{\tau e}\chi_e^{\;\;\mu} \mr \sim 3^*\+3'\+\tilde3\\
&\nonumber\\
&3'\x \tilde3^* \sim \ml h_{\mu\tau}\\ h_{\tau e}\\ h_{e\mu} \mr\!\x\!\ml \z^e_{\;\;\mu\tau}\\ \z^\mu_{\;\;\tau e}\\ \z^\tau_{\;\;e\mu} \mr = \left[ \begin{matrix} h_{\mu\tau}\z^e_{\;\;\mu\tau}\!+\!h_{\tau e}\z^\mu_{\;\;\tau e}\!+\!h_{e\mu}\z^\tau_{\;\;e\mu}\\ h_{\mu\tau}\z^e_{\;\;\mu\tau}\!+\w\,h_{\tau e}\z^\mu_{\;\;\tau e}\!+\w^*h_{e\mu}\z^\tau_{\;\;e\mu}\\ h_{\mu\tau}\z^e_{\;\;\mu\tau}\!+\w^*h_{\tau e}\z^\mu_{\;\;\tau e}\!+\w\,h_{e\mu}\z^\tau_{\;\;e\mu} \end{matrix}\right]\+\ml h_{e\mu}\z^e_{\;\;\mu\tau}\\ h_{\mu\tau}\z^\mu_{\;\;\tau e}\\ h_{\tau e}\z^\tau_{\;\;e\mu} \mr\+\ml h_{\tau e}\z^e_{\;\;\mu\tau}\\ h_{e\mu}\z^\mu_{\;\;\tau e}\\ h_{\mu\tau}\z^\tau_{\;\;e\mu} \mr \nonumber\\
&\qquad \sim \widetilde 1^*\+(\widetilde 1')^*\+(\widetilde 1'')^*\+\hat3\+\hat3^*
\end{align}
Next, we decompose the product of $\hat3$ with itself and with $\tilde3$:
\begin{align}
&\hat3\x\hat3 \sim \ml \chi_\tau^{\;\;\mu}\\ \chi_e^{\;\;\tau}\\ \chi_\mu^{\;\;e} \mr\!\x\!\ml \chi_\tau^{'\;\;\mu}\\ \chi_e^{'\;\tau}\\ \chi_\mu^{'\;e} \mr = \ml \chi_\tau^{\;\;\mu}\chi_\tau^{'\;\mu}\\ \chi_e^{\;\;\tau}\chi_e^{'\;\tau}\\ \chi_\mu^{\;\;e}\chi_\mu^{'\;e} \mr\+\ml \chi_\mu^{\;\;e}\chi_e^{'\;\tau}\\ \chi_\tau^{\;\;\mu}\chi_\mu^{'\;e}\\ \chi_e^{\;\;\tau}\chi_\tau^{\;'\mu} \mr\+\ml \chi_e^{\;\;\tau}\chi_\mu^{'\;e}\\ \chi_\mu^{\;\;e}\chi_\tau^{\;'\mu}\\ \chi_\tau^{\;\;\mu}\chi_e^{'\;\tau} \mr \sim (\hat3_1)^*\+(\hat3_2)^*\+(\hat3_3)^* \label{eq:3''x3''=...}\\
&\hat3\x \tilde3\sim \ml \chi_\tau^{\;\;\mu}\\ \chi_e^{\;\;\tau}\\ \chi_\mu^{\;\;e} \mr\!\x\!\ml \z_e^{\;\;\mu\tau}\\ \z_\mu^{\;\;\tau e}\\ \z_\tau^{\;\;e\mu} \mr = \ml \chi_\tau^{\;\;\mu}\z_\mu^{\;\;\tau e}\\ \chi_e^{\;\;\tau}\z_\tau^{\;\;e\mu}\\ \chi_\mu^{\;\;e}\z_e^{\;\;\mu\tau} \mr\+\ml \chi_e^{\;\;\tau}\z_\mu^{\;\;\tau e}\\ \chi_\mu^{\;\;e}\z_\tau^{\;\;e\mu}\\ \chi_\tau^{\;\;\mu}\z_e^{\;\;\mu\tau} \mr\+\ml \chi_\mu^{\;\;e}\z_\mu^{\;\;\tau e}\\ \chi_\tau^{\;\;\mu}\z_\tau^{\;\;e\mu}\\ \chi_e^{\;\;\tau}\z_e^{\;\;\mu\tau} \mr \sim 3^*\+3'\+\tilde3\label{eq:3''x3'''=...}
\end{align}
In contrast to the other triplet multiplications so far, the product of $\hat 3$ with its conjugate reduces to the nine singlets:
\begin{align}\label{eq:3''x(3'')^*=...}
\hat 3\x\hat3^*\sim  \ml \chi_\tau^{\;\;\mu}\\ \chi_e^{\;\;\tau}\\ \chi_\mu^{\;\;e} \mr\!\x\!\ml \chi_\mu^{'\;\tau}\\ \chi_\tau^{'\;e}\\ \chi_e^{'\;\mu} \mr &= \left[ \begin{matrix} \chi_\tau^{\;\;\mu}\chi_\mu^{'\;\tau}\!+\!\chi_e^{\;\;\tau}\chi_\tau^{'\;e}\!+\!\chi_\mu^{\;\;e}\chi_e^{'\;\mu}\\ \chi_\tau^{\;\;\mu}\chi_\mu^{'\;\tau}\!+\w\,\chi_e^{\;\;\tau}\chi_\tau^{'\;e}\!+\w^*\chi_\mu^{\;\;e}\chi_e^{'\;\mu}\\ \chi_\tau^{\;\;\mu}\chi_\mu^{'\;\tau}\!+\w^*\chi_e^{\;\;\tau}\chi_\tau^{'\;e}\!+\w\,\chi_\mu^{\;\;e}\chi_e^{'\;\mu} \end{matrix}\right] \begin{matrix} \!\leftarrow 1\\ \leftarrow 1'\\ \;\;\;\;\leftarrow (1')^*\end{matrix}\nonumber\\
&\+\left[ \begin{matrix} \chi_\tau^{\;\;\mu}\chi_\tau^{'\;e}\!+\!\chi_e^{\;\;\tau}\chi_e^{'\;\mu}\!+\!\chi_\mu^{\;\;e}\chi_\mu^{'\;\tau} \\  \chi_\tau^{\;\;\mu}\chi_\tau^{'\;e}\!+\w\,\chi_e^{\;\;\tau}\chi_e^{'\;\mu}\!+\w^*\chi_\mu^{\;\;e}\chi_\mu^{'\;\tau}\\  \chi_\tau^{\;\;\mu}\chi_\tau^{'\;e}\!+\w^*\chi_e^{\;\;\tau}\chi_e^{'\;\mu}\!+\w\,\chi_\mu^{\;\;e}\chi_\mu^{'\;\tau} \end{matrix} \right] \;\;\;\;\begin{matrix} \!\!\!\!\!\!\leftarrow \widetilde 1^*\\ \leftarrow (\widetilde 1'')^*\\ \!\leftarrow (\widetilde 1')^*\end{matrix} \nonumber\\
&\+\left[ \begin{matrix} \chi_\tau^{\;\;\mu}\chi_e^{'\;\mu}\!+\!\chi_e^{\;\;\tau}\chi_\mu^{'\;\tau}\!+\!\chi_\mu^{\;\;e}\chi_\tau^{'\;e}\\ \chi_\tau^{\;\;\mu}\chi_e^{'\;\mu}\!+\w\,\chi_e^{\;\;\tau}\chi_\mu^{'\;\tau}\!+\w^*\chi_\mu^{\;\;e}\chi_\tau^{'\;e}\\ \chi_\tau^{\;\;\mu}\chi_e^{'\;\mu}\!+\w^*\chi_e^{\;\;\tau}\chi_\mu^{'\;\tau}\!+\w\,\chi_\mu^{\;\;e}\chi_\tau^{'\;e} \end{matrix} \right]\;\;\;\; \begin{matrix} \!\!\leftarrow \widetilde 1\\ \!\leftarrow\widetilde 1'\\ \leftarrow \widetilde 1'' \end{matrix}
\end{align}
We also have:
\begin{align}\label{eq:3''x(3''')*=...}
&\hat 3\x \tilde 3^* \sim \ml \chi_\tau^{\;\;\mu}\\ \chi_e^{\;\;\tau}\\ \chi_\mu^{\;\;e} \mr\!\x\!\ml \z^e_{\;\;\mu\tau}\\ \z^\mu_{\;\;\tau e}\\ \z^\tau_{\;\;e\mu} \mr = \ml \chi_\tau^{\;\;\mu}\z^\tau_{\;\;e\mu}\\ \chi_e^{\;\;\tau}\z^e_{\;\;\mu\tau}\\ \chi_\mu^{\;\;e}\z^\mu_{\;\;\tau e} \mr\+\ml \chi_\mu^{\;\;e}\z^\tau_{\;\;e\mu}\\ \chi_\tau^{\;\;\mu}\z^e_{\;\;\mu\tau}\\ \chi_e^{\;\;\tau}\z^\mu_{\;\;\tau e} \mr\+\ml \chi_e^{\;\;\tau}\z^\tau_{\;\;e\mu}\\ \chi_\mu^{\;\;e}\z^e_{\;\;\mu\tau}\\ \chi_\tau^{\;\;\mu}\z^\mu_{\;\;\tau e} \mr \sim 3\+(3')^*\+\tilde 3^*
\end{align}
Lastly, we compute the product of $\tilde 3$ with itself and with its conjugate:
\begin{align}\label{eq:3'''x3'''=...}
\tilde 3\x \tilde3 &\sim \ml \z_e^{\;\;\mu\tau}\\ \z_\mu^{\;\;\tau e}\\ \z_\tau^{\;\;e\mu} \mr\!\x\!\ml \z_e^{'\;\mu\tau}\\ \z_\mu^{'\;\tau e}\\ \z_\tau^{'\;e\mu} \mr=\ml \z_e^{\;\;\mu\tau}\z_e^{'\;\mu\tau}\\ \z_\mu^{\;\;\tau e}\z_\mu^{'\;\tau e}\\ \z_\tau^{\;\;e\mu}\z_\tau^{'\;e\mu} \mr\+\ml \z_\mu^{\;\;\tau e}\z_\tau^{'\;e\mu}\\ \z_\tau^{\;\;e\mu}\z_e^{'\;\mu\tau}\\ \z_e^{\;\;\mu\tau}\z_\mu^{'\;\tau e} \mr\+\ml \z_\tau^{\;\;e\mu}\z_\mu^{'\;\tau e}\\ \z_e^{\;\;\mu\tau}\z_\tau^{'\;e\mu}\\ \z_\mu^{\;\;\tau e}\z_e^{'\;\mu\tau} \mr \sim \tilde 3^*\+3_1\+3_2
\end{align}
\begin{align}\label{eq:3'''x(3''')*=...}
\tilde 3\x \tilde 3^* &\sim \ml \z_e^{\;\;\mu\tau}\\ \z_\mu^{\;\;\tau e}\\ \z_\tau^{\;\;e\mu} \mr\!\x\! \ml \z'^e_{\;\;\mu\tau}\\ \z'^\mu_{\;\;\tau e}\\ \z'^\tau_{\;\;e\mu} \mr \!\!=\!\! \left[ \begin{matrix} \z_e^{\;\;\mu\tau}\z'^e_{\;\;\mu\tau}\!+\!\z_\mu^{\;\;\tau e}\z'^\mu_{\;\;\tau e}\!+\!\z_\tau^{\;\;e\mu}\z'^\tau_{\;\; e\mu} \\ \z_e^{\;\;\mu\tau}\z'^e_{\;\;\mu\tau}\!+\w\z_\mu^{\;\;\tau e}\z'^\mu_{\;\;\tau e}\!+\w^*\z_\tau^{\;\;e\mu}\z'^\tau_{\;\; e\mu}\\ \z_e^{\;\;\mu\tau}\z'^e_{\;\;\mu\tau}\!+\w^*\z_\mu^{\;\;\tau e}\z'^\mu_{\;\;\tau e}\!+\w\,\z_\tau^{\;\;e\mu}\z'^\tau_{\;\; e\mu}\end{matrix} \right]\!\!\+\!\!\ml \z_\mu^{\;\;\tau e}\z'^\tau_{\;\;e\mu}\\ \z_\tau^{\;\;e\mu}\z'^e_{\;\;\mu\tau}\\ \z_e^{\;\;\mu\tau}\z'^\mu_{\;\;\tau e} \mr\!\!\+\!\!\ml \z_\tau^{\;\;e\mu}\z'^\mu_{\;\;\tau e}\\ \z_e^{\;\;\mu\tau}\z'^\tau_{\;\;e\mu}\\ \z_\mu^{\;\;\tau e}\z'^e_{\;\;\mu\tau} \mr \nonumber\\
&\sim 1\+1'\+(1')^*\+\hat3\+\hat3^*
\end{align}
\\
Finally, observe that
\begin{equation}
9\times 1^2+8\times 3^2 = 81\qquad\text{ and }\qquad 9+8 = 17
\end{equation}
so that the theorem $\sum_{i\,=\,1}^{n_C}d_i^2 = 81$ is satisfied with $d_1 = 9$, $d_3 = 8$, all other $d_i = 0$, and $n_C = 17$.
\\\\
Therefore we have accounted for all of the irreducible representations of $\Sigma(81)$. The character table is given in Table~\ref{table:character table}.
\begin{table}[h]
\centering
\fbox{\begin{minipage}{18cm}
	\centering
	\tiny
	\renewcommand{\tabcolsep}{1mm}
	\begin{tabular}{c|c|*{18}{c|}}
          Class  &\# elts. & $\!\chi(1)\!$ & $\!\chi(1')\!$ & $\!\chi(\tilde 1)\!$ & $\!\chi(\tilde 1')\!$ & $\!\chi(\tilde 1'')\!$ & $\!\chi(3)\!$ & $\!\chi(3')\!$ & $\!\chi(\hat3)\!$ & $\!\chi(\tilde 3)\!$ & $\!\chi(1'^*)\!$ & $\!\chi(\tilde 1^*)\!$ & $\!\chi(\tilde 1'^*)\!$ & $\!\chi(\tilde 1''^*)\!$ & $\!\chi(3^*)\!$ & $\!\chi(3'^*)\!$ & $\!\chi(\hat3^*)\!$ & $\!\chi(\tilde 3^*)\!$ & $h$\\
\hline\hline
$C_I$ &$1$ & $1$&$1$&$1$&$1$&$1$&$3$&$3$&$3$&$3$&$1$&$1$&$1$&$1$&$3$&$3$&$3$&$3$&1\\ \hline
$C_\al$  &$3$ & $1$&$1$&$\w$&$\w$&$\w$&$2+\w$&$1+2\w$&$0$&$\w+2\w^*$&$1$&$\w^*$&$\w^*$&$\w^*$&$2+\w^*$&$1+2\w^*$&$0$&$\w^*+2\w$&3\\ \hline
$C_\al^*$  &$3$ & $1$&$1$&$\w^*$&$\w^*$&$\w^*$&$2+\w^*$&$1+2\w^*$&$0$&$\w^*+2\w$&$1$&$\w$&$\w$&$\w$&$2+\w$&$1+2\w$&$0$&$\w+2\w^*$&3\\ \hline
$C_{\al\beta}$  &$3$ & $1$&$1$&$\w^*$&$\w^*$&$\w^*$&$1+2\w$&$2\w+\w^*$&$0$&$3$&$1$&$\w$&$\w$&$\w$&$1+2\w^*$&$2\w^*+\w$&$0$&$3$&3\\ \hline
$C_{\al\beta}^*$  &$3$ & $1$&$1$&$\w$&$\w$&$\w$&$1+2\w^*$&$2\w^*+\w$&$0$&$3$&$1$&$\w^*$&$\w^*$&$\w^*$&$1+2\w$&$2\w+\w^*$&$0$&$3$&3\\ \hline
$D_\al^{\;\;\beta}$  &$3$ & $1$&$1$&$1$&$1$&$1$&$0$&$0$&$3\w$&$0$&$1$&$1$&$1$&$1$&$0$&$0$&$3\w^*$&$0$&3\\ \hline
$(D_\al^{\;\;\beta})^*$  &$3$ & $1$&$1$&$1$&$1$&$1$&$0$&$0$&$3\w^*$&$0$&$1$&$1$&$1$&$1$&$0$&$0$&$3\w$&$0$&3\\ \hline
$E_{\al\beta}^\g$  &$3$ & $1$&$1$&$\w$&$\w$&$\w$&$2\w+\w^*$&$2+\w^*$&$0$&$1+2\w$&$1$&$\w^*$&$\w^*$&$\w^*$&$2\w^*+\w$&$2+\w$&$0$&$1+2\w^*$&3\\ \hline
$(E_{\al\beta}^\g)^*$  &$3$ & $1$&$1$&$\w^*$&$\w^*$&$\w^*$&$2\w^*+\w$&$2+\w$&$0$&$1+2\w^*$&$1$&$\w$&$\w$&$\w$&$2\w+\w^*$&$2+\w^*$&$0$&$1+2\w$&3\\ \hline
$F$  &$1$ & $1$&$1$&$1$&$1$&$1$&$3\w$&$3\w^*$&$3$&$3\w^*$&$1$&$1$&$1$&$1$&$3\w^*$&$3\w$&$3$&$3\w$&3\\ \hline
$F^*$  &$1$ & $1$&$1$&$1$&$1$&$1$&$3\w^*$&$3\w$&$3$&$3\w$&$1$&$1$&$1$&$1$&$3\w$&$3\w^*$&$3$&$3\w^*$&3\\ \hline
$C_\al^c$  &$9$ & $1$&$\w$&$\w$&$\w^*$&$1$&$0$&$0$&$0$&$0$&$\w^*$&$\w^*$&$\w$&$1$&$0$&$0$&$0$&$0$&6\\ \hline
$C_\al^{c*}$  &$9$ & $1$&$\w^*$&$\w^*$&$\w$&$1$&$0$&$0$&$0$&$0$&$\w$&$\w$&$\w^*$&$1$&$0$&$0$&$0$&$0$&6\\ \hline
$D_\al^{c\;\beta}$  &$9$ & $1$&$\w$&$1$&$\w$&$\w^*$&$0$&$0$&$0$&$0$&$\w^*$&$1$&$\w^*$&$\w$&$0$&$0$&$0$&$0$&3\\ \hline
$C_\al^a$  &$9$ & $1$&$\w^*$&$\w$&$1$&$\w^*$&$0$&$0$&$0$&$0$&$\w$&$\w^*$&$1$&$\w$&$0$&$0$&$0$&$0$&6\\ \hline
$C_\al^{a*}$  &$9$ & $1$&$\w$&$\w^*$&$1$&$\w$&$0$&$0$&$0$&$0$&$\w^*$&$\w$&$1$&$\w^*$&$0$&$0$&$0$&$0$&6\\ \hline
$D_\al^{a\;\beta}$  &$9$ & $1$&$\w^*$&$1$&$\w^*$&$\w$&$0$&$0$&$0$&$0$&$\w$&$1$&$\w$&$\w^*$&$0$&$0$&$0$&$0$&3\\
\hline
	\end{tabular}
	\caption{Character table for $\Sigma(81)$. The phase $\w \equiv e^{\,i2\pi/3}$ satisfies $\w^2 = \w^*$, $\w^3 = 1$, and $1+\w+\w^* = 0$. The elements in each class can be inferred by comparing the index notation to the representations in Tables~\ref{table:generators on 1} and~\ref{table:generators on 3}. The last column (``$h$-value") denotes the minimum power to which the elements in each class must be raised in order to obtain the identity.}
	\label{table:character table}
\end{minipage}}
\end{table}
\pagebreak\\

\end{document}